\RequirePackage{fix-cm}
\documentclass[smallextended]{svjour3}       % onecolumn (second format)
\smartqed  % flush right qed marks, e.g., at end of proof
\usepackage{graphicx}
\usepackage{wrapfig}
\usepackage{hyperref}
\usepackage{alltt}
\usepackage{multirow}
\usepackage{graphicx}
\usepackage{color}
\usepackage[labelfont=bf,textfont={bf}]{caption}
\usepackage{subcaption}
\usepackage{pifont}% http://ctan.org/pkg/pifont
\usepackage{boxedminipage}
\usepackage{notoccite}
\usepackage{array}
\newcolumntype{C}[1]{>{\centering\let\newline\\\arraybackslash\hspace{0pt}}m{#1}}
\usepackage{enumerate}
\usepackage{booktabs}
\usepackage[framemethod=tikz]{mdframed}
\usepackage{lipsum}
\usetikzlibrary{shadows}
\usepackage{mathtools}
 \usepackage{paralist}
 \definecolor{light-gray}{gray}{0.80}
\usepackage{graphics}
\usepackage[shortlabels]{enumitem}
\usepackage{balance}
\usepackage{dblfloatfix}
\newmdenv[
  tikzsetting= {fill=light-gray},
  linewidth=1pt,
  roundcorner=0pt, 
  shadow=false
]{myshadowbox}
\usepackage{colortbl} 
\usepackage{blindtext, graphicx}
\usepackage{textcomp}
\usepackage{mathtools}
\usepackage[final]{listings}
\usepackage{balance}
\usepackage{picture}
\usepackage{relsize}
\usepackage{multicol}
\usepackage{soul}
\usepackage{enumitem}
\setitemize{noitemsep,topsep=0pt,parsep=0pt,partopsep=0pt,leftmargin=*}
\setenumerate{noitemsep,topsep=0pt,parsep=0pt,partopsep=0pt,leftmargin=*}
\usepackage{makecell}
\usepackage{bigstrut}

\newcommand{\fig}[1]{Figure~\ref{fig:#1}}

\newcommand{\tbl}[1]{Table~\ref{tbl:#1}}
\definecolor{comment_color}{rgb}{0.5, 0, 1}
\definecolor{steel}{rgb}{0.1, 0.3, 0.5} 

\newcommand{\bi}{\begin{itemize}}
\newcommand{\ei}{\end{itemize}}
\usepackage{verbatim}
\usepackage{algorithm}
\usepackage{algorithmicx}
\usepackage{algpseudocode}
\usepackage[export]{adjustbox}
\definecolor{LightCyan}{rgb}{0.88,1,1}
\definecolor{darkgray}{gray}{0.7}
\definecolor{Gray}{rgb}{0.88,1,1}
\definecolor{Gray}{gray}{0.85}
\definecolor{Blue}{RGB}{0,29,193}
\definecolor{MyDarkBlue}{rgb}{0,0.08,0.45} 
\definecolor{pink}{rgb}{.96,.72,.77}
\definecolor{lightergray}{rgb}{0.85, 0.85, 0.85}
\definecolor{darkgray}{rgb}{0.47, 0.47, 0.47}
\definecolor{lightestgray}{rgb}{0.95, 0.95, 0.95}
\definecolor{ao(english)}{rgb}{0.0, 0.5, 0.0}
\definecolor{beige}{rgb}{0.96, 0.96, 0.86}
\usepackage{tcolorbox}
\newtcolorbox{blockquote}{colback=black!5,boxrule=0.4pt,colframe=black,fonttitle=\bfseries}
\newtcolorbox{RQbox}{colback=red!5,boxrule=0.4pt,colframe=black,top=1pt,bottom=1pt,fonttitle=\bfseries}

\usepackage{cite}

\usepackage{listings}
\definecolor{MyDarkBlue}{rgb}{0,0.08,0.45} 
\begin{document}

\title{Revisiting Process versus Product  Metrics:  a Large Scale Analysis%\thanks{Grants or other notes
}
% \subtitle{Do you have a subtitle?\\ If so, write it here}


\author{Suvodeep Majumder         \and
        Pranav Mody \and Tim Menzies%etc.
}


\institute{S. Majumder \at
              Department of Computer Science, \\
              North Carolina State University,  Raleigh, USA \\
              \email{smajumd3@ncsu.edu}            %  \\
           \and
           P. Mody \at
              Department of Computer Science, \\
              North Carolina State University,   Raleigh, USA \\
              \email{prmody@ncsu.edu}
            \and 
            T. Menzies \at
              Department of Computer Science, \\
              North Carolina State University,  Raleigh, USA \\
              \email{tim@ieee.org}
}

\date{Received: date / Accepted: date}

\maketitle

\begin{abstract}
Numerous  methods can  build predictive models from software  data. However, what methods and conclusions should we endorse as we move from analytics in-the-small (dealing with a handful of projects) to analytics in-the-large (dealing with hundreds of projects)? 

To answer this question, we recheck prior small-scale results (about process versus product metrics for defect prediction and the granularity of metrics) using 722,471 commits from 700 Github projects. We find that some analytics in-the-small conclusions still hold  when scaling up to analytics in-the-large.  For example, like prior work,  we see that  process metrics are better predictors for defects than product metrics (best process/product-based learners respectively achieve  recalls of 98\%/44\% and AUCs of 95\%/54\%, median values). 

That said,  we warn that it is unwise to trust metric importance results from analytics in-the-small studies since those change dramatically when moving  to analytics in-the-large. Also, when reasoning in-the-large about hundreds of projects, it is better to use predictions from multiple models (since single model predictions can become confused and  exhibit a high variance).

% Apart from the above specific conclusions, our more general point is that the SE community now needs to  revisit many of the conclusions previously obtained via analytics in-the-small.


\end{abstract}

\section{Introduction}
\label{sec:intro}

There exist many automated software engineering techniques for building predictive models from software project data~\cite{ghotra15}. Such models are cost-effective methods for guiding developers on where to quickly find bugs~\cite{menzies2006data,ostrand04}.

Given that there are so many  techniques, the question naturally arises: which one should we use?
Software analytics is growing more complex and more ambitious with time.
A decade ago, a standard study in this field dealt with just 20 projects or less\footnote{For examples of such papers, see \tbl{lit}, later in this paper.}. Now we can access data on hundreds to thousands of projects. How does this change software analytics? What methods and conclusions should we endorse as we move from analytics in-the-small (which analyzes a small number of projects individually to report their findings) to analytics in-the-large (which analyzes hundreds of projects individually to report findings that are important across all or majority of the projects analyzed)\footnote{Note, here, when referring to analytics in-the-small and analytics in-the-large, we are not comparing findings from a local vs global approach. Rather we compare results and findings summarized from analyzing small number of projects vs results and findings summarized from analyzing large number of projects.}? So reproducing results and findings that were true for analytics in-the-small is of utmost importance with hundreds to thousands of projects. Such analytics in-the-large results will help the software engineering community to understand and adopt appropriate methods, beliefs, and  conclusions.

As part of this study, we revisited the Rahman et al. ICSE 2013  study {\em ``How, and why, process metrics are better''}~\cite{Ra13} and Kamei et al. ICSM 2010 study {\em ``Revisiting common bug prediction findings using effort-aware models''}~\cite{kamei2010revisiting}.
Both papers were analytics in-the-small study that used 12 and 3 projects, respectively to see if defect predictors worked best if they used:
\bi
    \item Product metrics, showing what was built; e.g., see \tbl{product}.
    \item Or process metrics, showing how code is changed; e.g., see \tbl{process};
\ei

These papers are worth revisiting since it is widely cited\footnote{232 and 179 citations respectively in Google Scholar, as of Sept 28, 2020.} and it addresses an important issue. Herbsleb argues convincingly that how groups organize themselves can be highly beneficial/detrimental to the process of writing code~\cite{Herbsleb14}. Hence, process factors can be highly informative about what parts of a codebase are buggy. In support of the Herbsleb hypothesis, prior studies have shown that, for defect prediction, process metrics significantly outperform product metrics~\cite{Lumpe12, Ra13, bird2009does}. 
Also, if we wish to learn general principles for software engineering that hold across multiple projects, it is better to use process metrics since:
\bi
    \item Process metrics are much simpler to collect and can be applied uniformly to software written in different languages.  
    \item Product metrics, on the other hand,  can be much harder to collect. For example, some  static code analysis requires expensive licenses, which need updating every time a new version of a language is released~\cite{Devanbu14}. Also, the collected value for these metrics may not translate between projects since those ranges can be highly specific.Lastly, product metrics tend to be far more verbose and hence time-consuming to collect. For example,  for 722,471 commits studied in this paper,  data collected required 500 days of CPU (using five machines, 16 cores, 7days). Our process metrics, on the other hand, were an order of magnitude faster to collect.\footnote{This is because process metrics can be calculate using the change history of a file. While calculating the product metrics, the tool needs to download the specific version of the file, then go through the actual code to gather the necessary statistics to calculate the actual metrics.}
\ei

\begin{table*}[t!]
\caption{List of product metrics used in this study}
\scriptsize
\begin{tabular}{c|l|c}
\rowcolor{gray!30} 
Type   & Metrics  & Count \\ 
File   & \begin{tabular}[l]{@{}l@{}}AvgCyclomatic, AvgCyclomaticModified, AvgCyclomaticStrict,\\ AvgEssential, AvgLine, AvgLineBlank, AvgLineCode, \\ AvgLineComment, CountDeclClassMethod, \\ CountDeclClassVariable, CountDeclInstanceMethod,  \\ CountDeclInstanceVariable, CountDeclMethod,  CountDeclMethodAll,\\ CountDeclMethodDefault, CountDeclMethodPrivate, \\ CountDeclMethodProtected, CountDeclMethodPublic,\\ CountLine, CountLineBlank,  CountLineCode, CountLineCodeDecl, \\ CountLineCodeExe, CountLineComment, CountSemicolon, CountStmt, \\ CountStmtDecl, CountStmtExe, MaxCyclomatic, \\ MaxCyclomaticModified, MaxCyclomaticStrict,MaxEssential, \\ RatioCommentToCode, SumCyclomatic, SumCyclomaticModified, \\SumCyclomaticStrict, SumEssential\end{tabular} & 37    \\ \hline
Class  & \begin{tabular}[c]{@{}l@{}}PercentLackOfCohesion, \\ PercentLackOfCohesionModified, MaxInheritanceTree, \\ CountClassDerived, CountClassCoupled, CountClassCoupledModified, \\ CountClassBase\end{tabular}                                                                                                                                                                                                                                                                                                                                                                                                                                                                                                                                                    & 7     \\ \hline
Method & MaxNesting                                                                                                          & 1      
\end{tabular}
\label{tbl:product}
\end{table*}

\begin{table}[!t]
\centering
\vspace{5mm}
\caption{List of process metrics used in this study}
\scriptsize
\begin{tabular}{r@{~:~}l}
% \hline
% \rowcolor[HTML]{C0C0C0}  
adev     &  Active Dev Count                                 \\  
age      &  Interval between the last and the current change \\  
ddev     &  Distinct Dev Count                               \\  
sctr  &  Distribution of modified code across each file   \\
exp      &  Experience of the committer                      \\  
la       &  Lines of code added                              \\  
ld       &  Lines of code deleted                            \\  
lt       &  Lines of code in a file before the change        \\  
minor    &  Minor Contributor Count                          \\  
nadev    &  Neighbor’s Active Dev Count                      \\  
ncomm    &  Neighbor’s Commit Count                          \\  
nd       &  Number of Directories                             \\  
nddev    &  Neighbor’s Distinct Dev Count                    \\  
ns       &  Number of Subsystems                             \\ 
nuc      &  Number of unique changes to the modified files   \\  
own      &  Owner’s Contributed Lines                        \\  
sexp     &  Developer experience on a subsystem              \\  
rexp     & Recent developer experience \\
\end{tabular}
\label{tbl:process}
\end{table}

Since product versus process metrics is such an important issue, we revisited the Rahman et al. and  Kamei et al. study. To check their conclusions, we ran an analytics in-the-large study that looked at 722,471 commits from 700 Github projects.
  
All in all, this paper explores eight hypotheses  using two widely used validation criteria. One is release-based (where given $R$ releases of the software, we trained on data from release 1 to $R-3$, then tested on release $R-2$, $R-1$, and $R$) and another is cross-validation based (where the data is randomly divided into $N$ stratified bins. Each bin, in turn, becomes the test set and a model is trained on the remaining bins.) After comparing conclusions seen in the prior analytics-in-the-small to the analytics-in-the-large, we find two cases where we disagree and six where we agree. So what is the value of a paper with  75\% agreement with prior work? We assert that this paper makes several important contributions:
\bi
    \item Firstly, in the two cases where we disagree, we very strongly disagree:
        \bi
            \item We find that the use of any learner is not appropriate for analytics-in-large. Our results suggest that any learner that generates a single model may get confused by all the intricacies  of data from multiple projects. On the other hand, {\em ensemble learners} (that make the conclusions by polling across many models) know how to generate good predictions from an extensive sample.
            \item Also, in terms of {\em what recommendations we would make} to improve software quality, we find that the conclusions achieved via  analytics-in-the-large are very different from those achieved via analytic-in-the-small. Later in this paper, we compare those two sets of conclusions. We will show that changes to software projects that make sense from analytics-in-the-small (after looking at any five projects)  can be wildly misleading since, once we get to analytics-in-the-large, a very different set of attributes is most effective
        \ei
    \item Secondly, in  the case where our conclusions are the same as prior work, we have  successfully completed a valuable step in the scientific process: i.e., reproduction of prior results. Current ACM guidelines\footnote{https://www.acm.org/publications/policies/artifact-review-and-badging-current} distinguish replication and reproduction as follows: the former uses artifacts from the  prior study while the latter does not. Our work is a {\em reproduction}\footnote{ To be clear: technically speaking, this paper is a {\em partial reproduction}  of  Rahman et al. or Kamei et al.  When we tried  their methodology, we found in some cases, our results needed a slightly different approach (see \S~\ref{sec:exp}).} since   we use ideas from the  Rahman et al. and Kamei et al. study, but none of their code or data. We would encourage more researchers to conduct and report more reproduction studies.
\ei 
  \noindent
Specifically, this paper asks eight research questions

\begin{RQbox}
\textbf{RQ 1:} For predicting defects, do methods that work in-the-small, also work in-the-large?
\end{RQbox}

In a result that agrees with  Rahman et al., we find that how we build code is more indicative of what bugs are introduced than what we build (i.e., process metrics make best defect predictions ).

\begin{RQbox}
\textbf{RQ 2:} Measured in terms of predication variability, do methods that works well in-the-small, also work at at-scale?
\end{RQbox}

Rahman et al. said  that it does not matter what learner is  used to build prediction models. {\underline{\bf We make the exact opposite conclusion.}} For  analytics-in-the-large, the more data we process, the more variance in that data. Hence, conclusions that rely on a single model get confused and exhibit significant variance in their predictions. To mitigate this problem, it is important to use learners that make conclusions by averaging over multiple models (i.e., ensemble Random Forests are far better for analytics than the Naive Bayes, Logistic Regression, or Support Vector Machines used in prior work).

\begin{RQbox}
\textbf{RQ 3:} Measured in terms of granularity, do same granularity that works well in-the-small, also work at at-scale?
\end{RQbox}

Kamei et al. said in their study that although the file-level prediction is better than package-level prediction when measured using Popt20, the difference is very little and we agree with this result. However, when measured via other evaluation measures, the difference is significantly different. Thus for analytics-in-the-large, when measured using other criteria, it is evident the granularity of the metrics matter and file-level prediction shows significantly better results than package-level prediction.

\begin{RQbox}
\textbf{RQ 4:} Measured in terms of stability, are process metrics more/less stable than code metrics, when measured at at-scale?    
\end{RQbox}
When measured in terms of stability of performance across the last 3 releases by using all other previous releases for training the model, our results agree with  Rahman et al. in all traditional evaluation criteria (i.e., recall, pf, precision). We find that the performance across the last 3 releases does not significantly differ in all evaluation criteria except for effort-aware  evaluation criteria Popt20. 

% \begin{RQbox}
% \textbf{RQ 5:} Measured in terms of portability, Are process metrics more/less portable than code metrics, when measured at at-scale?     
% \end{RQbox}

\begin{RQbox}
\textbf{RQ 5:} Measured in terms of stasis, Are process metrics more/less static than code metrics, when measured at at-scale?    
\end{RQbox}
In this result, we agree with Rahman et al.. We can see product metrics are significantly more correlated than process metrics. We measure this correlation in both release-based and JIT-based  settings. Although we can see process metrics have a significantly lower correlation than product metrics in both release-based and JIT-based  settings, the difference is lower in case of JIT-based  settings. Also, when lifting process metrics from file-level to package-level, as explored by Kamei et al., we can see a significant increase in correlation in case of process metrics. This can explain the drop in performance in package-level prediction.

\begin{RQbox}
\textbf{RQ 6:}  Measured in terms of stagnation, Do models built from different sets of metrics stagnate across releases, when measured at at-scale?      
\end{RQbox}
Rahman et al. warn that, when reasoning over multiple releases, models can   {\em stagnant}, i.e.,   fixate on old conclusions  and miss new ones. For example, if a defect occurs in the same file in release one and release two, and another defect appears in a new file in the second release, the model will catch the file as defective, which was defective in first release, but will miss the defect in the new file.
 
Here we measure the stagnation property of the models built using the metrics. Our results agree with Rahman et al.: we see a significantly higher correlation between the predicted probability and learned probability in the case of product metrics than process metrics. This signifies models built using product metrics tend to be stagnant.  

\begin{RQbox}
\textbf{RQ 7:} Do stagnant models (based on stagnant metrics) tend to predict recurringly defective entities?  
\end{RQbox}
In these results, we try to evaluate if models built with product and process metrics tend to predict recurrent defects. Our results concur with Rahman et al. and we see models built with product metrics tend to predict recurrent defects, while models built with process data do not suffer from this effect.

\begin{RQbox}
\textbf{RQ 8:} Measured in terms of metric importance, are metrics that seem important in-the-small, also  important when reasoning  in-the-large?    
\end{RQbox}

Numerous prior analytics in-the-small publications offer conclusions on the relative importance of different metrics. For example, \cite{Kamei10}, \cite{Gao11}, \cite{Moser08}, \cite{kondo2020impact}, \cite{d2010extensive} offer such conclusions after an analysis  of 1,1,3, 6,and 26 software project, respectively. Their conclusions are far more specific than process-vs-product; rather, these prior studies call our particular metrics are being most important for prediction.

Based on our analysis, we must now call into question any prior analytics in-the-small conclusions that assert that specific metrics are more important than any other (for defect prediction). We find that the relative importance of different \underline{{\bf metrics found via analytics in-the-small is not stable}}. Specifically, when we move to analytics in-the-large, we find very different rankings for metric importance. 

The rest of this paper is structured as follows. Some background and related work are discussed in section~\ref{sec:literature}. Our experimental methods are described in section~\ref{sec:experiment}. Data collection in section~\ref{sec:data} and learners used in this study in section~\ref{sec:learner}. Followed by the experimental setup in section~\ref{sec:exp} and evaluation criteria in section~\ref{sec:eval}. The results and answers to the research questions are presented in section~\ref{sec:results}. Which is followed by threats to validity in section~\ref{sec:threats}. Finally, the conclusion is provided in section~\ref{sec:conclusion}.

Note that all the scripts and data used in this analysis are available online at \url{https://github.com/Suvodeep90/Revisit\_process\_product} \footnote{Note to reviewers: Our data is so large we cannot place it in the Github repo. Zenodo.org will host our data. https://github.com/Suvodeep90/Revisit\_process\_product only contains a sample of our data. We will link that repository to link to data stored at Zenodo.org.}.

\section{Background and Related Work}
\label{sec:literature}

\subsection{Defect Prediction}
\label{sec:defect_prediction}

This section shows that software defect prediction is a (very) widely explored area with many application areas.
Specifically, in 2020, software defect prediction is now
a ``subroutine'' that enables much other research.

A defect in software is a failure or an error represented by incorrect, unexpected, or unintended behavior  of a system caused by an action taken by a developer. As today's software proliferates both in size and number, software testing for capturing those defects plays more and more crucial roles. During software development, the testing process  often has some resource limitations. For example, the effort associated with coordinated human effort across a large codebase can grow exponentially with the scale of the project \cite{fu2016tuning}.

It is common to match the quality assurance (QA) effort to the perceived criticality and bugginess of the code for managing resources efficiently. Since every decision is associated with a human and resource cost to the developer team, it is impractical and inefficient to distribute equal effort to every component in a software system\cite{briand1993developing}. Creating defect prediction models from either product metrics (like those from Table~\ref{tbl:product}) or process metrics (like those from Table~\ref{tbl:process}) is an efficient way to take a look at the incoming changes and focus on specific modules or files based on a suggestion from defect predictor.

% \begin{table}[!t]
% \centering
% \vspace{5mm}
% \caption{List of process metrics used in this study}
% \small
% \begin{tabular}{r@{~:~}l}
% % \hline
% % \rowcolor[HTML]{C0C0C0}  
% adev     &  Active Dev Count                                 \\  
% age      &  Interval between the last and the current change \\  
% ddev     &  Distinct Dev Count                               \\  
% sctr  &  Distribution of modified code across each file   \\
% exp      &  Experience of the committer                      \\  
% la       &  Lines of code added                              \\  
% ld       &  Lines of code deleted                            \\  
% lt       &  Lines of code in a file before the change        \\  
% minor    &  Minor Contributor Count                          \\  
% nadev    &  Neighbor’s Active Dev Count                      \\  
% ncomm    &  Neighbor’s Commit Count                          \\  
% nd       &  Number of Directories                             \\  
% nddev    &  Neighbor’s Distinct Dev Count                    \\  
% ns       &  Number of Subsystems                             \\ 
% nuc      &  Number of unique changes to the modified files   \\  
% own      &  Owner’s Contributed Lines                        \\  
% sexp     &  Developer experience on a subsystem              \\  
% rexp     & Recent developer experience \\
% \end{tabular}
% \label{tbl:process}
% \end{table}

Recent results show that software defect predictors  are also competitive widely-used  automatic methods.  Rahman et al. ~\cite{rahman2014comparing} compared (a) static code analysis tools FindBugs, Jlint, and PMD with (b) defect predictors (which they called ``statistical defect prediction'') built using logistic  regression. No significant differences in   cost-effectiveness were observed. Given this equivalence, it is significant to note that  defect prediction can be quickly adapted to new languages by building lightweight parsers to extract  product metrics or use common change information by mining git history to build process metrics. The same is not true for static code analyzers - these need extensive modification before they can be used in new languages. Because of this ease of use and its applicability to many programming languages, defect prediction has been   extended  in many ways, including:

\begin{enumerate}
    \item Application of defect prediction methods to locate code with security vulnerabilities~\cite{Shin2013}.
    \item Understanding the factors that lead to a greater likelihood of defects such as defect-prone software components using code metrics (e.g.,, ratio comment to code, cyclomatic complexity) \cite{menzies10dp,menzies07dp} or process metrics (e.g.,, recent activity).
    \item Predicting the location of defects so that appropriate resources may be allocated (e.g.,,~\cite{bird09reliabity})
    \item Using predictors to proactively fix  defects~\cite{arcuri2011practical}
    \item Studying defect prediction not only just release-level \cite{chen2018applications} but also change-level or just-in-time~\cite{commitguru}.  
    \item Exploring ``transfer learning'' where predictors from one project are applied to another~\cite{krishna2018bellwethers,nam18tse}.
    \item Assessing different learning methods for building predictors~\cite{ghotra15}. This has led to the development of hyper-parameter optimization and better data harvesting tools \cite{agrawal2018wrong,agrawal2018better}. 
\end{enumerate}

\begin{table}[]
\scriptsize
\begin{tabular}{|l|c|c|c|}
\hline
\rowcolor[HTML]{C0C0C0} 
Paper                                                                                                                                                                           & \begin{tabular}[c]{@{}c@{}}\# of \\ Datasets\end{tabular} & Year & Venue   \\ \hline
\begin{tabular}[c]{@{}l@{}}A validation of object-oriented design \\ metrics as quality indicators\end{tabular}                                                                 & 8                                                         & 1996 & TSE      \\ \hline
\begin{tabular}[c]{@{}l@{}}Predicting fault incidence using software \\ change history\end{tabular}                                                                             & 1                                                         & 2000 & TSE     \\ \hline
\begin{tabular}[c]{@{}l@{}}Empirical analysis of ck metrics for object-oriented \\ design complexity: Implications for software defects\end{tabular}                            & 1                                                         & 2003 & TSE      \\ \hline
\begin{tabular}[c]{@{}l@{}}Data mining static code attributes to learn defect \\ predictors\end{tabular}                                                                        & 8                                                         & 2006 & TSE     \\ \hline
\begin{tabular}[c]{@{}l@{}}Empirical analysis of object-oriented design metrics \\ for predicting high and low severity faults\end{tabular}                                     & 1                                                         & 2006 & TSE      \\ \hline
\begin{tabular}[c]{@{}l@{}}Is external code quality correlated with programming\\  experience or feelgoodfactor?\end{tabular}                                                   & 1                                                         & 2006 & XP         \\ \hline
Mining metrics to predict component failures                                                                                                                                    & 5                                                         & 2006 & TSE      \\ \hline
Predicting defects for eclipse                                                                                                                                                  & 1                                                         & 2007 & ICSE     \\ \hline
\begin{tabular}[c]{@{}l@{}}The effects of over and under sampling on fault-prone \\ module detection\end{tabular}                                                               & 1                                                         & 2007 & ESEM       \\ \hline
\begin{tabular}[c]{@{}l@{}}Using software dependencies and churn metrics to \\ predict field failures: An empirical case study\end{tabular}                                     & 1                                                         & 2007 & ESEM       \\ \hline
\begin{tabular}[c]{@{}l@{}}A comparative analysis of the efficiency of change \\ metrics and static code attributes for defect prediction\end{tabular}                          & 1                                                         & 2008 & ICSE       \\ \hline
\begin{tabular}[c]{@{}l@{}}Benchmarking  models for  defect   prediction\end{tabular}                                                                    & 10                                                        & 2008 & TSE         \\ \hline
\begin{tabular}[c]{@{}l@{}}Do too many cooks spoil the broth? using the number \\ of developers to enhance defect prediction models\end{tabular}                                & 2                                                         & 2008 & EMSE       \\ \hline
Implications of ceiling effects in defect predictors                                                                                                                            & 12                                                        & 2008 & IPSE      \\ \hline
\begin{tabular}[c]{@{}l@{}}An investigation of the relationships between lines \\ of code and defects\end{tabular}                                                              & 1                                                         & 2009 & ICSE     \\ \hline
\begin{tabular}[c]{@{}l@{}}On the relative value of cross-company and \\ within-company data for defect prediction\end{tabular}                                                 & 6                                                         & 2009 & EMSE       \\ \hline
\begin{tabular}[c]{@{}l@{}}Cross-project defect prediction: a large scale \\ experiment on data vs. domain vs. process\end{tabular}                                             & 7                                                         & 2009 & FSE       \\ \hline
\begin{tabular}[c]{@{}l@{}}A systematic and comprehensive investigation of \\ methods to build and evaluate fault prediction model\end{tabular}                                 & 1                                                         & 2010 & JSS       \\ \hline
An analysis of developer metrics for fault prediction                                                                                                                           & 1                                                         & 2010 & PROMISE     \\ \hline
Change bursts as defect predictors                                                                                                                                              & 1                                                         & 2010 & ISSRE       \\ \hline
Predicting faults in high assurance software                                                                                                                                    & 15                                                        & 2010 & HASE       \\ \hline
\begin{tabular}[c]{@{}l@{}}Revisiting Common Bug Prediction Findings Using \\ Effort-Aware Models\end{tabular}                                                                  & 3                                                         & 2010 & ICSM    \\ \hline
Bugcache for inspections: hit or miss                                                                                                                                           & 5                                                         & 2011 & FSE       \\ \hline
\begin{tabular}[c]{@{}l@{}}Don’t touch my code! examining the effects of \\ ownership on software quality\end{tabular}                                                          & 2                                                         & 2011 & FSE       \\ \hline
\begin{tabular}[c]{@{}l@{}}Ownership, experience and defects: a \\ fine-grained study of authorship\end{tabular}                                                                & 4                                                         & 2011 & ICSE       \\ \hline
\begin{tabular}[c]{@{}l@{}}Using coding-based ensemble learning to \\ improve software defect prediction\end{tabular}                                                           & 14                                                        & 2012 & SMC         \\ \hline
\begin{tabular}[c]{@{}l@{}}Transfer learning for cross-company \\ software defect prediction\end{tabular}                                                                       & 6                                                         & 2012 & IST     \\ \hline
\begin{tabular}[c]{@{}l@{}}Recalling the “imprecision” of \\ cross-project defect prediction.\end{tabular}                                                                      & 9                                                         & 2012 & FSE        \\ \hline
Method-level bug prediction                                                                                                                                                     & 21                                                        & 2012 & ESEM       \\ \hline
How, and why, process metrics are better                                                                                                                                        & 12                                                        & 2013 & ICSE       \\ \hline
\begin{tabular}[c]{@{}l@{}}Using class imbalance learning for software \\ defect prediction\end{tabular}                                                                        & 10                                                        & 2013 & TR       \\ \hline
Sample Size vs. Bias in Defect Prediction                                                                                                                                       & 12                                                        & 2013 & FSE        \\ \hline
Predicting Bugs Using Antipatterns                                                                                                                                              & 2                                                         & 2013 & ICSME     \\ \hline
\begin{tabular}[c]{@{}l@{}}Empirical study of the classification performance \\ of learners on imbalanced noisy software quality data\end{tabular}                              & 1                                                         & 2014 & IS       \\ \hline
\begin{tabular}[c]{@{}l@{}}Learning to rank relevant files for bug reports \\ using domain knowledge\end{tabular}                                                               & 6                                                         & 2014 & FSE         \\ \hline
\begin{tabular}[c]{@{}l@{}}Which process metrics can significantly improve \\ defect prediction models? an empirical study\end{tabular}                                         & 11                                                        & 2014 & MSR         \\ \hline
\begin{tabular}[c]{@{}l@{}}The impact of mislabelling on the performance \\ and interpretation  of defect prediction models\end{tabular}                                        & 5                                                         & 2015 & ICSE      \\ \hline
\begin{tabular}[c]{@{}l@{}}Developer Micro Interaction Metrics for \\ Software Defect Prediction\end{tabular}                                                                   & 6                                                         & 2016 & TSE     \\ \hline
\begin{tabular}[c]{@{}l@{}}Hydra: Massively compositional model for \\ cross-project defect prediction\end{tabular}                                                             & 10                                                        & 2016 & TSE  \\ \hline
\begin{tabular}[c]{@{}l@{}}Supervised vs Unsupervised Models: A Holistic \\ Look at Effort-Aware Just-in-Time Defect Prediction\end{tabular}                                    & 6                                                         & 2017 & ICSME     \\ \hline
\begin{tabular}[c]{@{}l@{}}Empirical analysis of change metrics for software \\ fault prediction\end{tabular}                                                                   & 1                                                         & 2018 & CEE       \\ \hline
\begin{tabular}[c]{@{}l@{}}CLEVER: Combining Code Metrics with Clone \\ Detection for Just-In-Time Fault Prevention and \\ Resolution in Large Industrial Projects\end{tabular} & 1                                                         & 2018 & MSR       \\ \hline
Fine-grained just-in-time defect prediction                                                                                                                                     & 10                                                        & 2019 & IST         \\ \hline
\begin{tabular}[c]{@{}l@{}}Mining  defects: Should we consider   affected releases?\end{tabular}                                                                       & 6                                                         & 2019 & ICSE       \\ \hline
\end{tabular}
\caption{Number of data sets explored in recent papers 
at prominent venues that experiment with process and/or product metrics.}
\label{tbl:lit}
\end{table}

% There has been many studies on defect prediction to identify which learner to use, what types of attributes to learn from. Many of these studies report conflicting results~\cite{sun2012using, seiffert2009improving, menzies2008implications,seliya2010predicting, khoshgoftaar2002cost}. The reasons include differing data sets, experimental design and performance measures along with differing parameterization approaches for the classifiers. This makes it very hard to determine what to conclude and what advice to give practitioners seeking to predict -defect-prone software components. This study tries to explore 4 different different learners widely used in defect prediction on a much larger dataset of 770 projects on both product and process metrics. This is to showcase if one has an advantage over others and to verify if they predict the same modules as defective or not?

\subsection{Process vs Product}
\label{sec:process_vs_product}

Defect prediction models are built using various machine learning classification methods such as Random Forest, Support Vector Machine, Naive Bayes, Logistic Regression ~\cite{tantithamthavorn2016automated,zhang2016cross,jacob2015improved,zhang2007predicting,ibrahim2017software,wang2013using,krishna2018bellwethers,sun2012using,menzies2008implications,seiffert2014empirical,seliya2010predicting,ghotra2015revisiting,zhang2017data,he2012investigation,nam2013transfer,pan2010domain} etc. All these methods input project metrics and output a model that can make predictions. Fenton et al.~\cite{fenton2000software} say that a ``metric'' is an attempt to measure some internal or external characteristic and can broadly be classified into {\em product} (specification, design, code-related) or {\em process} (constructing specification, detailed design related). The metrics are    computed either through parsing the codes (such as modules, files, classes or methods) to extract product (code) metrics or by inspecting the change history by parsing the revision history of files to extract process (change) metrics. 

In September  2020, we conducted the following literature review to understand the current thinking on the process and product metrics. Starting with  Rahman et al.~\cite{rahman2013and} and Kamei et al.~\cite{kamei2010revisiting}, we used Google Scholar to trace citations forward and backward-looking for papers that offered experiments on the process or product metrics for defect prediction or that suggested why certain process or product metrics are better for defect prediction. This gave us a list of 76 papers. Following the advice of Mathew et al.~\cite{mathew2017trends}, we examined:
\bi
    \item Highly cited papers, i.e., those with at least ten cites per year.
    \item Papers from senior SE venues, i.e., those listed at ``\href{https://scholar.google.com/citations?view_op=top_venues&hl=en&vq=eng_softwaresystems}{Google Scholar Metrics Software Systems}''.
\ei
Next, using our domain expertise, we augmented that list of papers we considered important or highly influential papers that focus on the benefits of using process or/and product metrics that were not included in the above two criteria). This leads to the 45 papers that are listed in  Table~\ref{tbl:lit}.

% Finally, one last paper was added since, as far as we could tell, it was the first to discuss this issue in the context of analytics.

\begin{figure}[!b]
\centering
    \includegraphics[width=.6\linewidth]{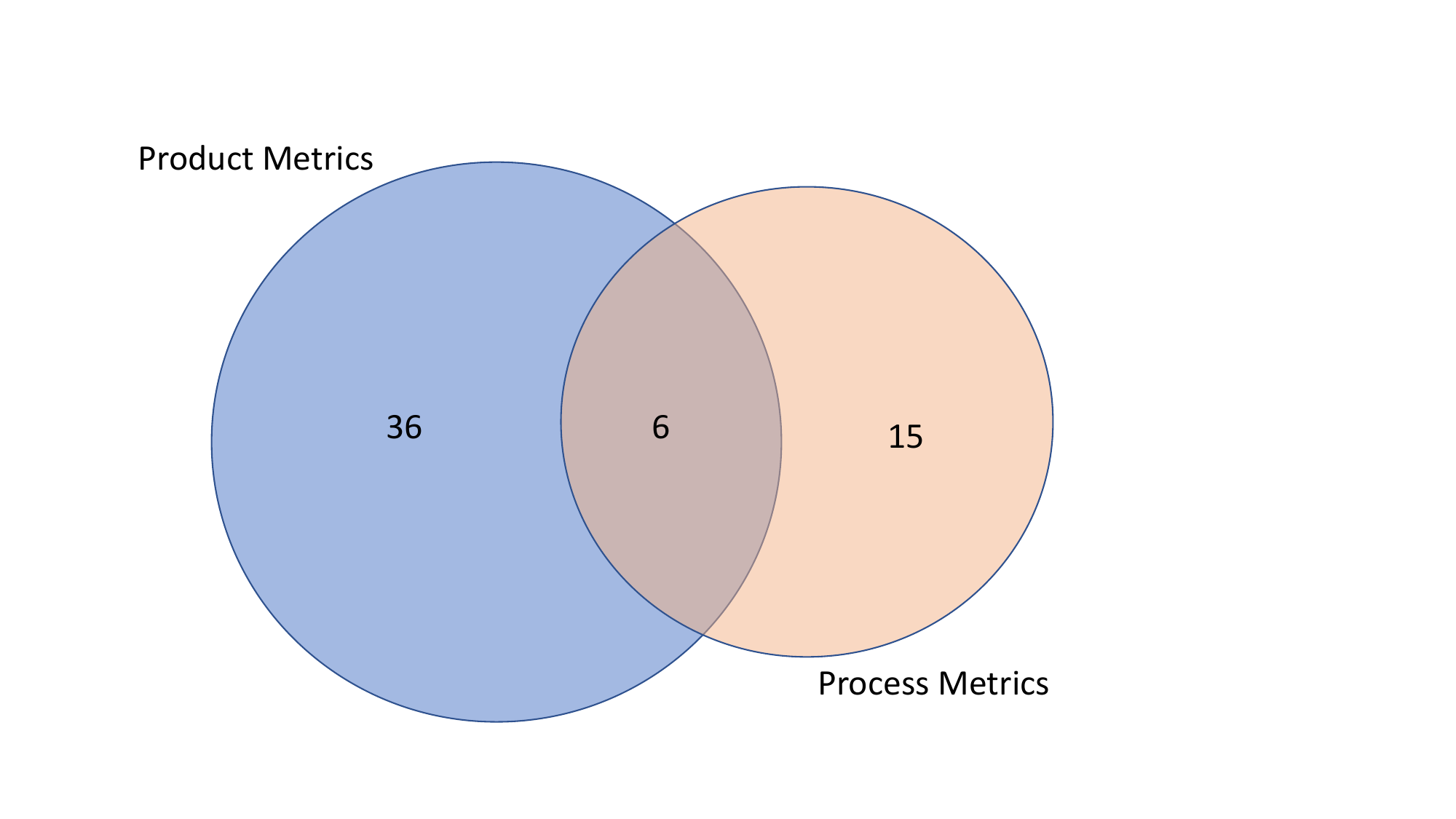}
    \caption{Number of papers exploring the benefits of the process and product metrics for defect prediction. The papers in the  intersection are~\cite{rahman2013and,moser2008comparative,graves2000predicting,arisholm2010systematic,kamei2010revisiting, giger2012method} explore and compare both process and product metrics. Note that prior to this EMSE paper, prior work that looked at the process and product metrics explored analytics-in-the-small.} 
    \label{fig:lit}
\end{figure}

Within this set of papers, we observe that studies on product metrics are more common than on process metrics (and very few papers experimentally compare both product and process metrics: see  Figure~\ref{fig:lit}). The product metrics community~\cite{wang2013using, sun2012using, menzies2008implications, seiffert2014empirical, seliya2010predicting, kamei2007effects, menzies2006data, zimmermann2007predicting, turhan2009relative, zimmermann2009cross, xia2016hydra} argues that   many kinds of metrics indicate which code modules are buggy:
\bi
    \item For example, for lines of code, it is usually argued that  large files can be hard to comprehend and change  (and thus are more likely to have bugs);
    \item For another example, for design complexity, it is often argued that the more complex a design of   code, the    harder it is to change and improve that code (and thus are more likely to have bugs).
\ei

On the other hand, the process  metrics community~\cite{bird2011don,nagappan2007using,rahman2011ownership,rahman2011bugcache,weyuker2008too,madeyski2015process,choudhary2018empirical, nayrolles2018clever, tantithamthavorn2015impact, pascarella2019fine, rahman2013sample, huang2017supervised, ye2014learning} explore many process  metrics, including (a)~developer's experience; and (b)~how many developers worked on certain file (and, it is argued, many developers working on a single file is much more susceptible to defects); and (c)~how long it has been since the last change (and, it is argued, a file which is changed frequently may be an indicator for bugs).
 
The rest of this section lists prominent results from the Figure~\ref{fig:lit} survey. From the product metrics community, Zimmermann et al.~\cite{zimmermann2007predicting}, in their study on Eclipse project using file and package-level data, showed complexity-based product metrics are much better in predicting defective files. Zhang et al.~\cite{zhang2009investigation}, in their experiments, showed  that lines of code-related metrics are good predictors of software defects using NASA datasets. In another study using product metrics, Zhou et al.~\cite{zhou2010ability} analyzed a combination of ten object-oriented software metrics related to  complexity to conclude that  size metrics were a much better indicator of defects. A similar study  by Zhou and Leung et al.~\cite{zhou2006empirical} evaluated the importance of individual metrics and indicated that while CBO, WMC, RFC, and LCOM metrics are useful metrics for fault prediction, but DIT is not useful using NASA datasets. Menzies et al.~\cite{menzies2006data}, in their study regarding static code metrics for defect prediction, found product metrics are very effective in finding defects. Basili et al.~\cite{basili1996validation}, in their work, showed object-oriented ck metrics appeared to be useful in predicting class fault-proneness, which was later confirmed by Subramanyam and Krishnan et al.~\cite{subramanyam2003empirical}. Nagappan et al.~\cite{nagappan2006mining}, in their study, reached a similar conclusion as Menzies et al.~\cite{menzies2006data}, but concluded, ``However, there is no single set of complexity metrics that could act as a universally best defect predictor''. 

In other studies related to process metrics, Nagappan et al.~\cite{nagappan2010change} emphasized the importance of change bursts as a predictor for software defects on Windows Vista dataset. They achieved a precision and recall value at 90\% in this study and achieved a precision of 74.4\% and recall at 88.0\% in another study on  Windows Server 2003 datasets. In another study by Matsumoto et al.~\cite{matsumoto2010analysis} investigated the effect of developer-related metrics on defect prediction. They showed improved performance using these metrics and proved module that is revised by more developers tends to contain more faults. Similarly, Schröte et al.~\cite{madeyski2006external}, in their study, showed a high correlation between the number of developers for a file and the number of defects in the respective file. 

As to  the six papers that compare process versus product methods:
\bi
    \item Four of these papers argue that process metrics are best. Rahman et al.~\cite{rahman2013and} found process metrics perform much better than product metrics in both within-project and cross-project defect prediction settings. Their study also showed product metrics do not evolve much over time and that they  are much more static. Hence, they say, product metrics are not  good predictors for defects. Similar conclusions (about the superiority of process metrics) are offered by  Moser et al.~\cite{moser2008comparative}, Giger et al.~\cite{giger2012method},  and  Graves et al.~\cite{graves2000predicting}.
    \item Only one  paper argues that both process and product metrics perform similarly. Arisholm et al.~\cite{arisholm2010systematic} found one project where both process and product metrics perform similarly. 
    \item Only one  paper argues that the combination of process and product metrics is better at predicting deefects. Kamei et al.~\cite{kamei2010revisiting} found 5 out of 9 versions of 3 projects where  combination of process and product metrics perform better than just using process metrics and 9 out of 9 cases they are better than just using product metrics. 
\ei

Of these papers,    Moser et al.~\cite{moser2008comparative}, Arisholm et al.~\cite{arisholm2010systematic}, Kamei et al.~\cite{kamei2010revisiting}, Rahman et al.~\cite{rahman2013and}, Graves et al.~\cite{graves2000predicting} and Giger et al.~\cite{giger2012method} based their conclusions on 1,1,3,12,15,21 projects (respectively). That is to say, these are all analytics in-the-small studies. The rest of this paper checks their conclusions using analytics in-the-large.

\section{Methods}
\label{sec:experiment}

This section describes our methods for   comparatively evaluating process versus product metrics using analytics in-the-large. 

\subsection{Data Collection}
\label{sec:data}
To collect data, we search Github for Java projects from   different software development domains.  Although Github stores millions of projects, many of these are trivially very small, not maintained, or are not about -software development projects.  To filter projects, we used the standard Github ``sanity checks'' recommended in the literature ~\cite{perils, curating, agrawal2018we}:

\begin{itemize}
    \item {\textit{{Collaboration}}: refers to the  number of pull requests. This is indicative of how many other peripheral developers work on this project. We required all  projects to have at least one pull request. This will prove the repository is a part of distributed development model where others have forked/created a branch on this repository to make independent changes and submitted those changes to the main repository to be merged with the main branch. We also validated and remove  any project where all pull requests are submitted by same developers by checking unique ids of pull request submitter.}
    \item {\textit{{Commits}}: The project must contain more than 20 commits as recommended in the literature. Commits in a  Github repository represent the amount of activity in the project. More than 75\% of the projects found in Github have less than 20; thus 20 is a good number for this filtering criteria.}
    \item {\textit{{Duration}}: The project must contain software development activity of at least 50 weeks. Kalliamvakou et al. show in their paper the 75\% of the project are active for less than 14 weeks; thus 50 weeks as a minimum duration for the filtering criteria is used as suggested by other researchers. }
    \item {\textit{{Issues}}: The project must contain more than 10 issues as recommended in the literature.}
    \item {\textit{{Releases}}: The project must contain at least 4 releases. This is because the release-based validation strategy used in this study requires 3 test releases and at least one training release.}
    \item {\textit{{Personal Purpose}}: The project must not be used and maintained by one person. The project must have at least eight contributors as suggested by other researchers.}
    \item {\textit{{Software Development}}: The project must only be a placeholder for software development source code.}
    \item {\textit{{Defective Commits}}: The project must have at least 10 defective commits with defects on Java files. This is because the SMOTE algorithm that we are using for balancing the datasets requires at least 10 examples of the minority class.}
    \item {\textit{Forked Project}}:  The project must not be a forked project from the original repository.This is to remove any potential duplicity and remove any project from the study that is not the project's main branch. We used the Github API to check for the ``Forked'' flag, and we removed any project which is flagged as yes.
\end{itemize}

We started with 8023 Github projects from various domains collected using Github search API. After applying the sanity checks mentioned above,  we selected 700 projects. The Data Statistics section of Table~\ref{tbl:stats} shows the median and IQR of each of the filtering criteria for the selected projects. For this research, we collected file-level process metrics and file-level product metrics to answer our research questions (RQ1, RQ3-RQ8) as suggested by Rahman et al.~\cite{rahman2013and}. We also followed the suggested aggregation process used by Kamei et al.~\cite{kamei2010revisiting} in their paper to calculate the package-level metrics by lifting the file-level metrics to the package-level to investigate and answer RQ2.

This data was extracted once and stored as pickle files in the following four steps:

\begin{table}[!t]
\scriptsize
\begin{tabular}{rrrrrr}
\rowcolor[HTML]{C0C0C0} 
\multicolumn{6}{c}{\cellcolor[HTML]{C0C0C0}Product Metrics}                                                                                                                                         \\
\rowcolor[HTML]{C0C0C0} 
\multicolumn{1}{c}{\cellcolor[HTML]{C0C0C0}Metric Name} & Median & \multicolumn{1}{c|}{\cellcolor[HTML]{C0C0C0}IQR} & \multicolumn{1}{c}{\cellcolor[HTML]{C0C0C0}Metric Name}   & Median  & IQR     \\
AvgCyclomatic                                           & 1      & \multicolumn{1}{r|}{1}                           & CountLine                                                 & 75.5    & 150     \\
AvgCyclomaticModified                                   & 1      & \multicolumn{1}{r|}{1}                           & CountLineBlank                                            & 10.5    & 20      \\
AvgCyclomaticStrict                                     & 1      & \multicolumn{1}{r|}{1}                           & CountLineCode                                             & 53      & 105     \\
AvgEssential                                            & 1      & \multicolumn{1}{r|}{0}                           & CountLineCodeDecl                                         & 18      & 32      \\
AvgLine                                                 & 9      & \multicolumn{1}{r|}{10}                          & CountLineCodeExe                                          & 29      & 66      \\
AvgLineBlank                                            & 0      & \multicolumn{1}{r|}{1}                           & CountLineComment                                          & 5       & 18      \\
AvgLineCode                                             & 7      & \multicolumn{1}{r|}{8}                           & CountSemicolon                                            & 24      & 52      \\
AvgLineComment                                          & 0      & \multicolumn{1}{r|}{1}                           & CountStmt                                                 & 35      & 72.3    \\
CountClassBase                                          & 1      & \multicolumn{1}{r|}{0}                           & CountStmtDecl                                             & 15      & 28      \\
CountClassCoupled                                       & 3      & \multicolumn{1}{r|}{4}                           & CountStmtExe                                              & 19      & 43.8    \\
CountClassCoupledModified                               & 3      & \multicolumn{1}{r|}{4}                           & MaxCyclomatic                                             & 3       & 4       \\
CountClassDerived                                       & 0      & \multicolumn{1}{r|}{0}                           & MaxCyclomaticModified                                     & 2       & 4       \\
CountDeclClassMethod                                    & 0      & \multicolumn{1}{r|}{0}                           & MaxCyclomaticStrict                                       & 3       & 5       \\
CountDeclClassVariable                                  & 0      & \multicolumn{1}{r|}{1}                           & MaxEssential                                              & 1       & 0       \\
CountDeclInstanceMethod                                 & 4      & \multicolumn{1}{r|}{7.5}                         & MaxInheritanceTree                                        & 2       & 1       \\
CountDeclInstanceVariable                               & 1      & \multicolumn{1}{r|}{4}                           & MaxNesting                                                & 1       & 2       \\
CountDeclMethod                                         & 5      & \multicolumn{1}{r|}{9}                           & \%LackOfCohesion                                     & 33      & 71      \\
CountDeclMethodAll                                      & 7      & \multicolumn{1}{r|}{12.5}                        & \%LackOfCohesionModified                             & 19      & 62      \\
CountDeclMethodDefault                                  & 0      & \multicolumn{1}{r|}{0}                           & RatioCommentToCode                                        & 0.1     & 0.2     \\
CountDeclMethodPrivate                                  & 0      & \multicolumn{1}{r|}{1}                           & SumCyclomatic                                             & 8       & 17      \\
CountDeclMethodProtected                                & 0      & \multicolumn{1}{r|}{0}                           & SumCyclomaticModified                                     & 8       & 17      \\
CountDeclMethodPublic                                   & 3      & \multicolumn{1}{r|}{6}                           & SumCyclomaticStrict                                       & 9       & 18      \\
SumEssential                                            & 6      & \multicolumn{1}{r|}{11}                          &                                                           &         &         \\ \hline
\rowcolor[HTML]{C0C0C0} 
\multicolumn{3}{c|}{\cellcolor[HTML]{C0C0C0}Process Metrics}                                                        & \multicolumn{3}{c}{\cellcolor[HTML]{C0C0C0}Data Statistics}                   \\
\rowcolor[HTML]{C0C0C0} 
\multicolumn{1}{c}{\cellcolor[HTML]{C0C0C0}Metric Name} & Median & \multicolumn{1}{c|}{\cellcolor[HTML]{C0C0C0}IQR} & \multicolumn{1}{c}{\cellcolor[HTML]{C0C0C0}Data Property} & Median  & IQR     \\
la                                                      & 14     & \multicolumn{1}{r|}{38.9}                        & Defect Ratio                                              & 37.60\% & 20.60\% \\
ld                                                      & 7.9    & \multicolumn{1}{r|}{12.2}                        & Lines of Code                                             & 82K     & 200K    \\
lt                                                      & 92     & \multicolumn{1}{r|}{121.8}                       & Number of Files                                           & 171     & 358     \\
age                                                     & 28.8   & \multicolumn{1}{r|}{35.1}                        & Number of Developers                                      & 31      & 34      \\
ddev                                                    & 2.4    & \multicolumn{1}{r|}{1.2}                         & Number of PRs.                                            & 55      & 101    \\
nuc                                                     & 5.8    & \multicolumn{1}{r|}{2.7}                         & Number of Commits                                         & 217     & 379    \\
own                                                     & 0.9    & \multicolumn{1}{r|}{0.1}                         & Duration                                                  & 186(W)  & 191(W) \\
minor                                                   & 0.2    & \multicolumn{1}{r|}{0.4}                         & Number of Releases                                        & 20      & 32     \\
ndev                                                    & 22.6   & \multicolumn{1}{r|}{22.1}                        & Number of Defective  Commits                              & 77      & 139     \\
ncomm                                                   & 71.1   & \multicolumn{1}{r|}{49.5}                        & Number of Issues                                          & 46      & 67      \\
adev                                                    & 6.1    & \multicolumn{1}{r|}{2.9}                         & Number of unique PR submitter                             & 5       & 6      \\
nadev                                                   & 71.1   & \multicolumn{1}{r|}{49.5}                        &                                                           &         &         \\
avg\_nddev                                              & 2      & \multicolumn{1}{r|}{1.8}                         &                                                           &         &         \\
avg\_nadev                                              & 7      & \multicolumn{1}{r|}{5.2}                         &                                                           &         &         \\
avg\_ncomm                                              & 7      & \multicolumn{1}{r|}{5.2}                         &                                                           &         &         \\
ns                                                      & 1      & \multicolumn{1}{r|}{0}                           &                                                           &         &         \\
exp                                                     & 348.8  & \multicolumn{1}{r|}{172.7}                       &                                                           &         &         \\
sexp                                                    & 145.7  & \multicolumn{1}{r|}{70}                          &                                                           &         &         \\
rexp                                                    & 2.5    & \multicolumn{1}{r|}{3.4}                         &                                                           &         &         \\
nd                                                      & 1      & \multicolumn{1}{r|}{0}                           &                                                           &         &         \\
sctr                                                    & -0.2   & \multicolumn{1}{r|}{0.1}                         &                                                           &         &        
\end{tabular}
\vspace{5mm}
\caption{Statistical median and IQR values for the metrics used in this study (IQR denotes the (75-25)th percentile range).}
\label{tbl:stats}
\end{table}

\begin{enumerate}
    \item We collected 21 process metrics (following the definition either from commit\_guru or from the definitions shared by Rahman et al.) for each file in each commit by  extracting the commit history of the project, then analyzing each commit for our metrics. We used a modified version of Commit\_Guru~\cite{rosen2015commit} code for this purpose, where instead of aggregating file-specific metric values for a commit, we store metric values for each file. We create objects for each new file we encounter and keep track of details (i.e., developer who worked on the file, LOCs added, modified, deleted by each developer, etc.) that we need to calculate. We also keep track of files modified together to calculate co-commit-based metrics.  After collecting the 21 metrics as mentioned in Table~\ref{tbl:stats} for each project, it is stored as a pickle file to be used for prediction.
    \item Secondly, we use Commit\_Guru~\cite{rosen2015commit} code to identify buginducing  and bugfixing commits. This process involves identifying bugfixing commits using a keyword\footnote{The keywords used are - \textit{bug, fix, error, issue, crash, problem, fail, defect and patch}. These keywords are taken used by Rosen et al. in their commit\_guru~\cite{rosen2015commit} paper.} based search. Using these commits, the process uses the  commit\_guru's SZZ algorithm~\cite{williams2008szz, rosen2015commit} to find commits that were responsible for introducing those changes and marking them as buginducing \footnote{From this point onwards, we will denote the commit which has bugs in them as a ``buginducing ''}. This process is performed on all commits throughout the life cycle of the project. Note here for a buginducing, each file that is labeled as a buggy file (buginducing ) will have another instance of the same file, which is non-buggy (bugfixing). If a file has been fixed multiple times throughout the project history, it will have multiple instances in the dataset.
    \item  Thirdly, we used Github tag API to collect the release information for each of the projects. We use the release number, release date information supplied from the API to group commits into releases and thus dividing each project into multiple releases for each of the metrics. Note here we refer to a release number as the tags provided by the contributors of the repository, not by Github. Thus we apply regular expressions to match the release number to either ``X.X.X.X'' or ``X.X.X'' format. Here for a tag to be considered as a release, it needs to be different in the section before the third dot.
    \item Finally, we used the Understand from Scitools\footnote{http://www.scitools.com/} to extract the 45 product metrics used in this study. Understand has a command-line interface to analyze project codes and generate metrics from that. We use the data collected from the first 2 steps to generate a list of commits and their corresponding files, along with class labels for defective and non-defective files. Next, we download the project codes from Github, then used the {\tt git commit} information to move the git head to the corresponding commit to match the code for that commit. Understand uses this snapshot of the code to analyze the metrics for each file and store the data in temporary storage. We do this for all commits throughout the project history. To ensure for every analyzed commit, we only consider the files which were changed, and we only keep files which was changed as part of that commit.
    Here we also added the class labels to the metrics. To only mark files that were defective,  we use commit Ids along with file names to add labels.  After the last step is done, the 45 product metrics collected for each project are stored in a separate file to answer the research questions for this study.
\end{enumerate}
Note that steps one and two required  2 days (on a single 16 cores machine), while step four required 7 days (on 5 machines with 16 cores) of computation, respectively. The data collected in this way are summarized in Table~\ref{tbl:stats}.

\subsection{Learners}
\label{sec:learner}
In this section, we briefly explain the four classification methods we have used for this study. We selected the following based on a prominent paper by Ghotra et al.'s~\cite{ghotra2015revisiting}. Also, all these learners  are  widely used in the software engineering community. For all the following models, we use the implementation from Scikit-Learn\footnote{https://scikit-learn.org/stable/index.html}. We applied Differential Evolution (DE) as a hyperparameter optimization~\cite{Tantithamthavorn18} to tune the models discussed here. However, as shown below, the performance of the Random Forest model with default parameters was so promising that we applied hyperparameter optimization on 3 of the models except for Random Forest. 

\subsubsection{\textbf{Support Vector Machine}} This is a discriminative classifier, which tries to create a hyper-plane between classes by projecting the data to a higher dimension using kernel tricks~\cite{ryu2016value,cao2018improved,tomar2015comparison,menzies2018500+}. The model learns the separating hyper-plane from the training data and classifies test data based on which side the example resides.

\subsubsection{\textbf{Naive Bayes}} This is a probabilistic model, widely used in software engineering community~\cite{wang2013using,sun2012using,menzies2008implications,seiffert2014empirical,seliya2010predicting}, that finds patterns in the training dataset and builds predictive models. This learner assumes all the   variables used for prediction are not correlated, identically distributed. This classifier uses Bayes rules to build the classifier. When predicting for test data, the model uses the distribution learned from training data to calculate the probability of the test example belonging  to each class and report the class with maximum probability. 

\subsubsection{\textbf{Logistic Regression}} This is a statistical predictive analysis method similar to linear regression but uses a logistic function to make predictions. Given 2 classes Y=(0 or 1) and a metric vector $X = {x_1,x_2,....,x_n}$, the learner first learns coefficients of each metrics vector to best match the training data. When predicting for test examples, it uses the metrics vectors of the test example and the coefficients learned from training data to make the prediction using  a logistic function. Logistic regression is widely used in defect prediction~\cite{ghotra2015revisiting,zhang2017data,he2012investigation,nam2013transfer,pan2010domain}. 

\subsubsection{\textbf{Random Forest}} This is a type of ensemble learning method, which consists of multiple classification decision trees built on random metrics and bootstrapped samples selected from the training data.  Test examples are classified by each decision tree in the Random Forest and then the final classification decision  is decided using a majority voting. Random forest is widely used in software engineering domain~\cite{tantithamthavorn2016automated,zhang2016cross,jacob2015improved,zhang2007predicting,ibrahim2017software,wang2013using,krishna2018bellwethers} and has proven to be effective in defect prediction. 

Later in this paper, the following distinction will become very significant. Of the four learners we apply, Random Forests make their conclusion via a majority vote across {\em  multiple models} while all the other learners build and apply a {\em single model}.

% This study uses scikit learn implementation of Random Forest, which build 100 decision tree with number of features equal to $\sqrt{n\_features}$.

\subsection{Differential Evolution (DE)}
\label{sec:hpo}

In this section, we explain the hyper-parameter optimizer used in this study to fine-tune  an ML model's parameters. There are several parameters for each ML model, which decide how an ML model learns to discriminate between desirable and undesirable outcomes. These parameters of the models can greatly affect the performance of the models. In this study, we used Differential Evolution (DE) as the hyper-parameter optimized as has been widely used in software engineering and machine learning community~\cite{tantithamthavorn2018impact, agrawal2018better, xia2018hyperparameter, onan2016multiobjective}.  DE is a stochastic population-based optimization algorithm~\cite{storn1997differential}. DE starts with a frontier of randomly generated candidate solutions. For example, when exploring tuning, each member of the frontier would be a different possible set of control settings for (say) an Support Vector Machine.

After initializing this frontier, a new candidate solution is generated by extrapolating by some factor $f$ between other items on the frontier. Such extrapolations are performed for all attributes at probability {\em cf}. If the candidate is better than one item of the frontier, then the candidate replaces the frontier item. The search then repeats for the remaining frontier items. For the definition of ``better``, this study uses F1-score; i.e.,  ``better'' means maximizing  the objective score of the model-based F1 Score. This process is repeated for {\em lives} number of repeated traversals of the frontier. For full details of DE,  see Figure~\ref{fig:pseudo_DE}. As per Storn's advice~\cite{storn1997differential}, we use \[f=0.75, \mathit{cf}=0.3, \mathit{lives}=60\]

\begin{figure}[!t]
\centering
\begin{lstlisting}[mathescape, frame=none, numbers=right]
def DE( n=10, cf=0.3, f=0.7):  # default settings
    frontier = sets of guesses (n=10)
    best = frontier.1 # any value at all
    lives = 1
    while(lives$--$ > 0): 
      tmp = empty
      for i = 1 to $|$frontier$|$: # size of frontier
         old = frontier$_i$
         x,y,z = any three from frontier, picked at random
         new= copy(old)
         for j = 1 to $|$new$|$: # for all attributes
           if rand() < cf    # at probability cf...
              new.j = $x.j + f*(z.j - y.j)$  # ...change item j
         # end for
         new  = new if better(new,old) else old
         tmp$_i$ = new
         if better(new,best) then
            best = new
            lives++ # enable one more generation
         end                  
      # end for
      frontier = tmp
    # end while
    return best
\end{lstlisting}
\caption{Differential Evolution based on Storn\textquotesingle s DE optimizer.}
\label{fig:pseudo_DE}
\vspace{-0.3cm}
\end{figure}

Out of the 4 learners, as mentioned in Section~\ref{sec:learner}, we have tuned 3 learners (a)~Logistic Regression, (b)~Naive Bayes, and (c)~Support Vector Machine. We did not include the Random Forest learner as it was already reporting near-perfect results for most performance measures. The parameters tuned in DE for each learner are - 

\begin{itemize}
    \item \textbf{Logistic Regression:} (a)~{\em penalty}: Used to specify the norm used in the penalization, (b)~$C$: Inverse of regularization strength, (c)~{\em solver}: Algorithm to use in the optimization problem, and (d)~{\em max\_iter}: Maximum number of iterations taken for the solvers to converge.
    \item\textbf{ Naive Bayes:} (a)~{\em var\_smoothing}: Portion of the largest variance of all features that are added to variances for calculation stability.
    \item \textbf{Support Vector Machine:} (a)~$C$: Regularization parameter, (b)~{\em gamma}: Kernel coefficient, (c)~{\em kernel}: Specifies the kernel type to be used in the algorithm, and (d)~{\em coef0}: Independent term in kernel function. 
\end{itemize}

\subsection{Experimental Framework}
\label{sec:exp}

\begin{figure*}[!t]
    \includegraphics[width=\linewidth]{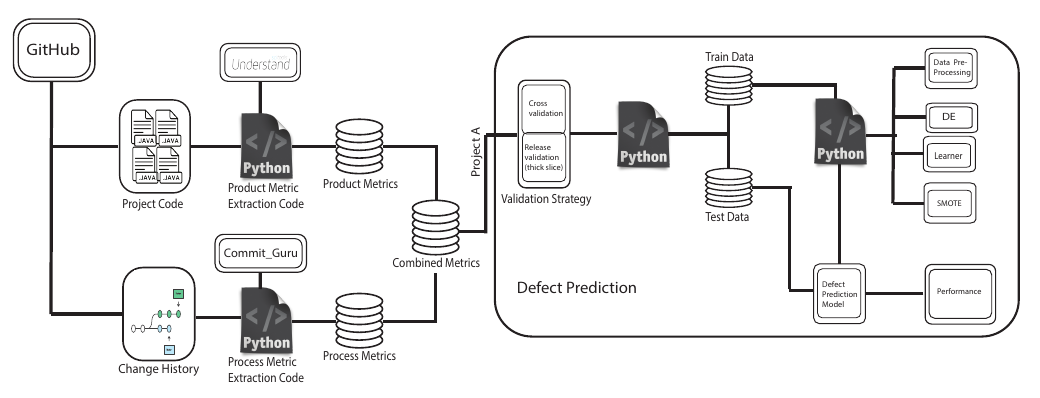}
    \caption{Framework for this analysis.} 
    \label{fig:framework}
\end{figure*}

Figure~\ref{fig:framework} illustrates  our  experimental  rig. For each of our 700 selected Java projects, we first use the project's revision history to collect file-level change metrics,  along with class labels (defective and non-defective commits). Then, using information from the process metrics, we use Understand's command-line interface to collect and filter the product metrics. Next, we join the two metrics to create a combined metrics set for each project. 

Using the evaluation strategy mentioned above, the data is divided into train, validation and test sets. The data is then filtered depending on metrics we are interested in (i.e., process, product, or combined) and pre-processed (i.e., data normalization, filtering/imputing missing values, etc.). After pre-processing and metric filtering is completed, the data is processed using SMOTE algorithm to handle data imbalance.  As described by Chawla et al.~\cite{chawla2002smote}, SMOTE is  useful for re-sampling training data such that a learner can find rare target classes. For more details in SMOTE, see~\cite{chawla2002smote, agarwal17}. Note one technical detail: when applying SMOTE, it is important that it is {\em not } applied to the validation or test data since data mining models need to be tested on the kinds of data they might actually see in practice.

Finally, we select one learner from four and it is applied to the training set to build a model. If hyperparameter optimization is to be performed, then the model is tuned using the validation data. Finally, the model is tested using the test data. As to how we generate our train/test sets, we report results from two methods: 
\begin{enumerate}
    \item {\em release-based}  
    \item {\em cross-validation} 
\end{enumerate}

Both these methods are defined below. We use both methods since (a)~other software analytics papers use {\em cross-validation} while  (b)~{\em release-based } is  the evaluation procedure of Rahman et al. As we shall see, these  two methods offer very similar results so debates about the merits of one approach to the other are something  of a moot point.  
But by reporting on results from both methods, it is more likely that other researchers will be able to compare their results against ours.

In a {\em cross-validation study}, we select all the files collected using the process described in Section~\ref{sec:data}. This includes the files that were labeled as buggy and non-buggy (this can include multiple copies of the same file if it was committed multiple times) throughout the project history. This data for each project is sorted randomly $M$ times. Then for each time, the data is divided into $N$ stratified bins. Each bin, in turn, becomes the test set and the remaining data is further divided into training and validation sets. For this study, we used $M=N=5$.
    
An alternative to cross-validation is a {\em release-based approach} such as the one used by Rahman et al. Here; given $R$ releases of the software, we divide all the data into $R$ parts. Then we trained on data from release 1 to $R-3$, then tested on release $R-2$, $R-1$, and $R$. This temporal approach has the advantage that the future data never appears in the training data.

\subsection{Evaluation Criteria}
\label{sec:eval}

In this section, we introduce the following 6 evaluation measures used in this study to evaluate the performance of machine learning models. Based on the results of the defect predictor, humans read the code in order of what the learner says is most defective. During that process, they find true negative, false negative, false positive, and true positive  (labeled TN, FN, FP, TP, respectively) reports from the learner.

\textbf{Recall:} This is the proportion of inspected defective changes among all the actual defective changes; i.e., TP/(TP+FN).   Recall is used in  many previous studies~\cite{kamei2012large, Tu18Tuning, yang2016effort, yang2017tlel, xia2016collective, yang2015deep}.   When recall is maximal, we are finding all the target class items. Hence we say that {\em larger} recalls are {\em better}.
    
\textbf{Precision:} This is the proportion of inspected defective changes among all the inspected changes; i.e.,  TP/(TP+FP). When precision is maximal, all the reports of defect modules are actually buggy (so the users waste no time looking at results that do not matter to them). Hence we say that {\em larger} precision is {\em better}.
    
\textbf{Pf:} This is the proportion of all suggested defective changes that are not actual defective changes divided by everything that is not actually  defective; i.e., FP/(FP+TN). A high {\em pf} suggests developers will be inspecting code that is not buggy.  Hence we say that {\em smaller} false alarms are {\em better}. 
    
    % \item \textbf{G-Score:} This is a harmonic mean between recall and (1-pf), defined as $\frac{recall*(1-pf)}{(recall+pf)}$.
    
\textbf{Popt20:} A good defect predictor lets programmers find the most bugs after reading the least amount of code\cite{Arisholm:2006}. {\bf Popt20} models that criteria. First, we divide the test data into (a) those that are predicted to be defective and (b)~those that are not. Second, we sorted the sets (a,b) on LOC. Third, we returned the test in the order sorted (a) followed by sorted (b). Within that sort, we then report the percent of actual bugs  found by inspecting the first 20\% of the code (measured in terms of LOC). We say that {\em larger} Popt20 values are {\em better}.
    
\textbf{IFA:}  Parnin and Orso~\cite{parnin2011automated} warn that developers will ignore the suggestions of static code analysis tools if those tools offer too many false alarms before reporting something of interest. Other researchers echo that concern~\cite{parnin2011automated,kochhar2016practitioners,xia2016automated}. {\bf IFA}  counts the number of  initial false alarms encountered before we find the first defect.  We say that {\em smaller} IFA values are {\em better}.
    
\textbf{AUC\_ROC:} This is the area under the curve for receiver operating characteristic. This is designated by a curve between true positive and false positive rates and created by varying the thresholds for defects between 0 and 1. This creates a  curve between (0,0) and (1,1), where a model with random guess will yield a value of 0.5 by connecting (0,0) and (1,1) with a straight line. A model with better performance will yield a higher value with a more convex curve in the upper left part. Hence we say that {\em larger} AUC values are {\em better}.

\subsection{Statistical Tests}
\label{stats}

When comparing the results of different models in this study, we used a statistical significance test and an effect size test:
\bi
    \item Significance test is useful for detecting if two populations differ merely by random noise. 
    \item Effect sizes are useful for checking that two populations differ by more than just a trivial amount.
\ei
For the significance test,  we use the Scott-Knott procedure  recommended at TSE'13~\cite{mittas2013ranking} and ICSE'15~\cite{ghotra2015revisiting}. This technique recursively bi-clusters a sorted set of numbers. If any two clusters are statistically indistinguishable, Scott-Knott reports them both as belonging to the same ``rank''.

To generate these ranks, Scott-Knott first looks for a break in the sequence that maximizes the expected values in the difference in the means before and after the break. More specifically,  it  splits $l$ values into sub-lists $m$ and $n$ to maximize the expected value of differences  in the observed performances before and after divisions. e.g.,, list $l,m$ and $n$ of size $ls,ms$ and, $ns$ where $l=m\cup n$, Scott-Knott divides the sequence at the break that maximizes:

\begin{equation}
    E(\Delta)=\frac{ms}{ls}\times abs(m.\mu - l.\mu)^2  + \frac{ns}{ls}\times abs(n.\mu - l.\mu)^2
\end{equation}

Scott-Knott then applies some statistical hypothesis test $H$ to check if $m$ and $n$ are significantly different. If so, Scott-Knott then recurses on each division. For this study, our hypothesis test $H$ was a conjunction of the A12 effect size test (endorsed by \cite{arcuri2011practical})  and non-parametric bootstrap sampling \cite{efron94}, i.e., our Scott-Knott divided the data if {\em both} bootstrapping and an effect size test agreed that the division was statistically significant (90\% confidence) and not a ``small'' effect ($A12 \ge 0.6$).

\section{RESULTS}
\label{sec:results}

\begin{RQbox}
    \textbf{RQ 1:} For predicting defects, do methods that work in-the-small, also work in-the-large? 
\end{RQbox} 

\begin{figure}[!b]
\centering
    \includegraphics[width=0.8\linewidth]{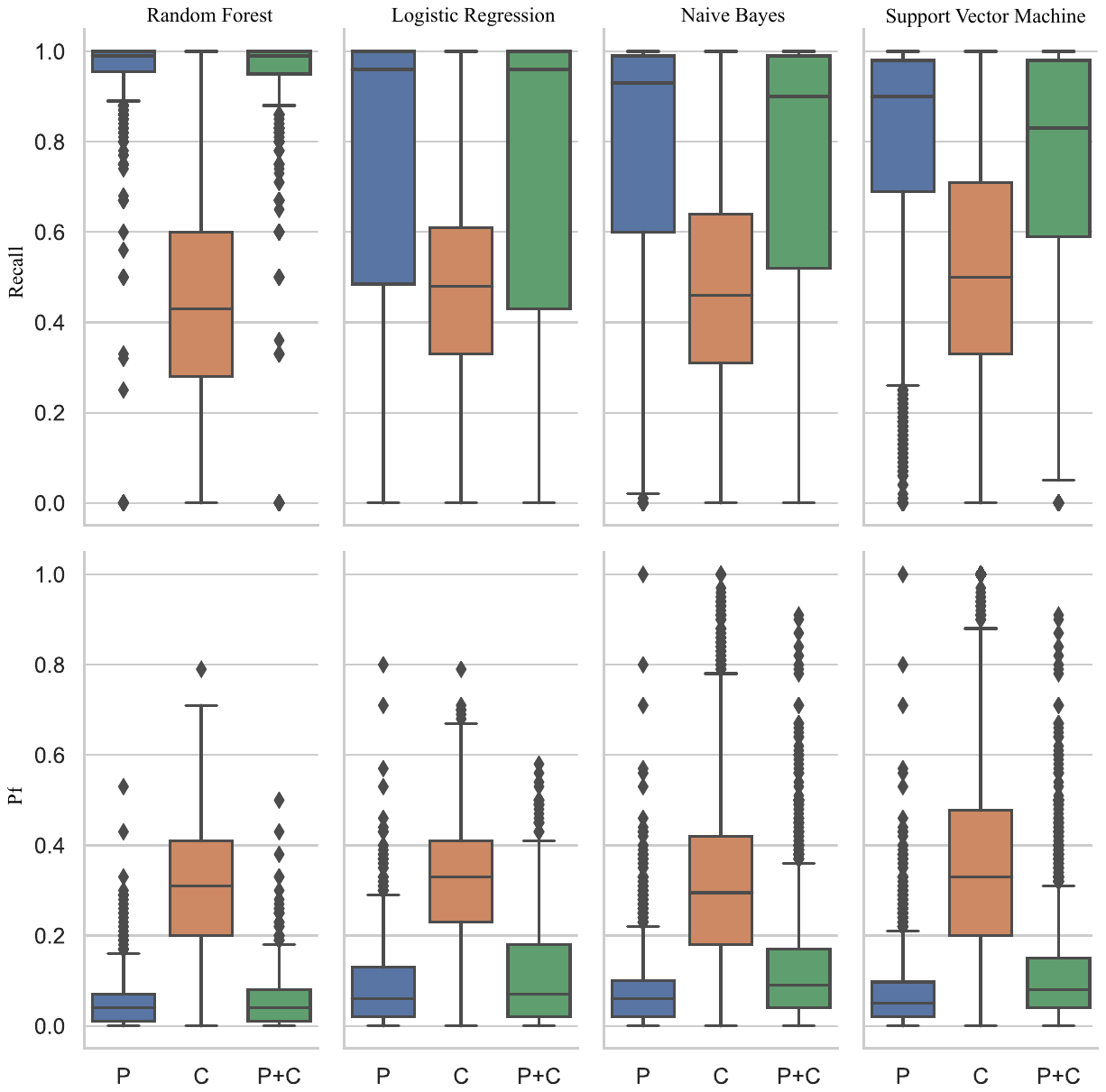}
    \caption{Cross-validation recall and false alarm results for   Process(P), Product(C) and, Combined (P+C) metrics.
    The vertical box plots in these charts run from min to max while the thick boxes highlight the 25,50,75th percentile.
    Each box  plot  is  built  using  700  Github  projects,  where  each  data  point  is  the(a) median result from 5-fold cross-validation repeated 5 times. } 
    \label{fig:learner_performance_1}
\end{figure}

\begin{figure}[!b]
\centering
    \includegraphics[width=0.8\linewidth]{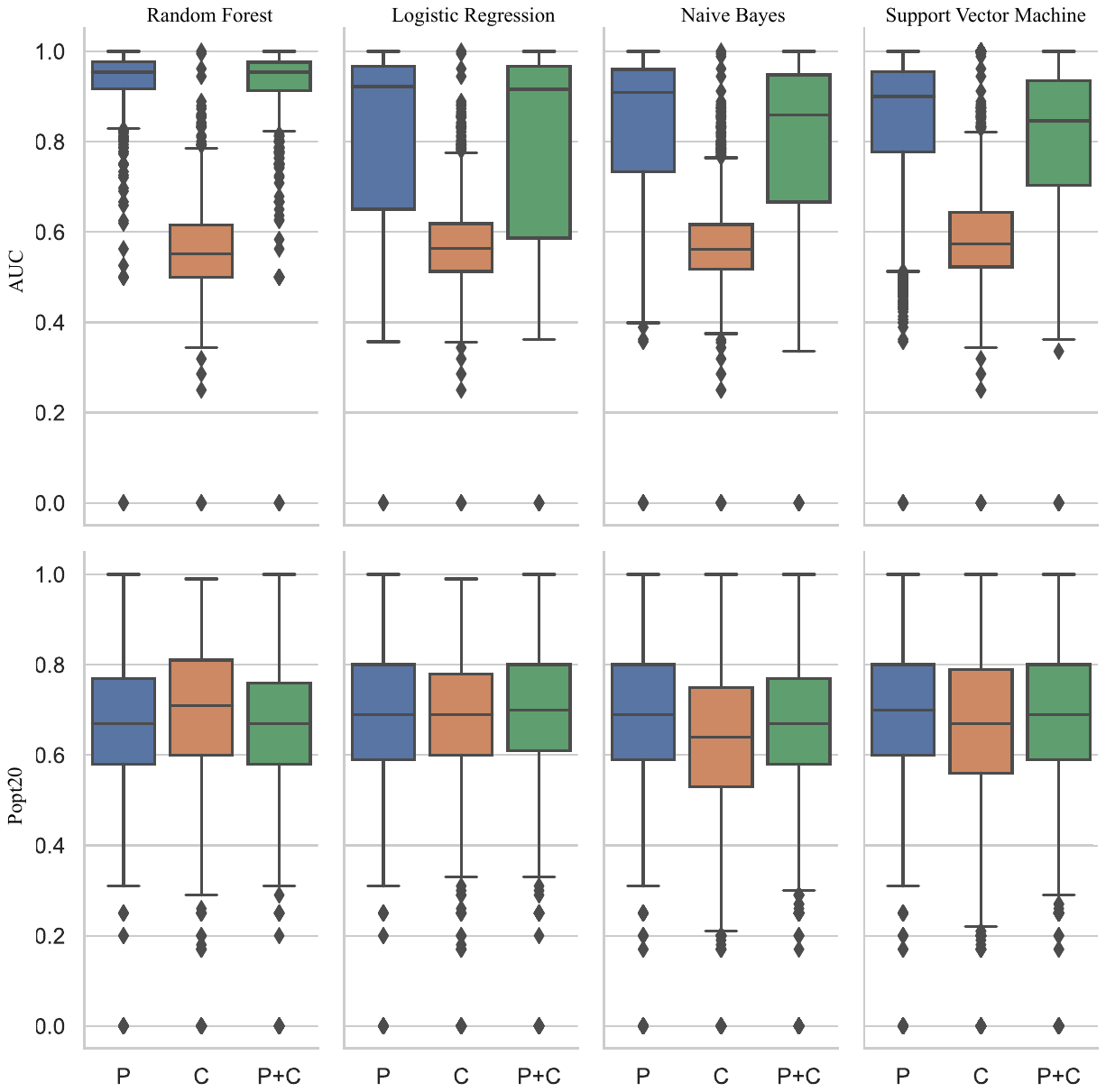}
    \caption{Cross-validation AUC and Popt20 results for  Process(P), Product(C), and Combined (P+C) metrics.
    Same format as \fig{learner_performance_1}.} 
    \label{fig:learner_performance_2}
\end{figure}

\begin{figure}[!b]
\centering
    \includegraphics[width=0.8\linewidth]{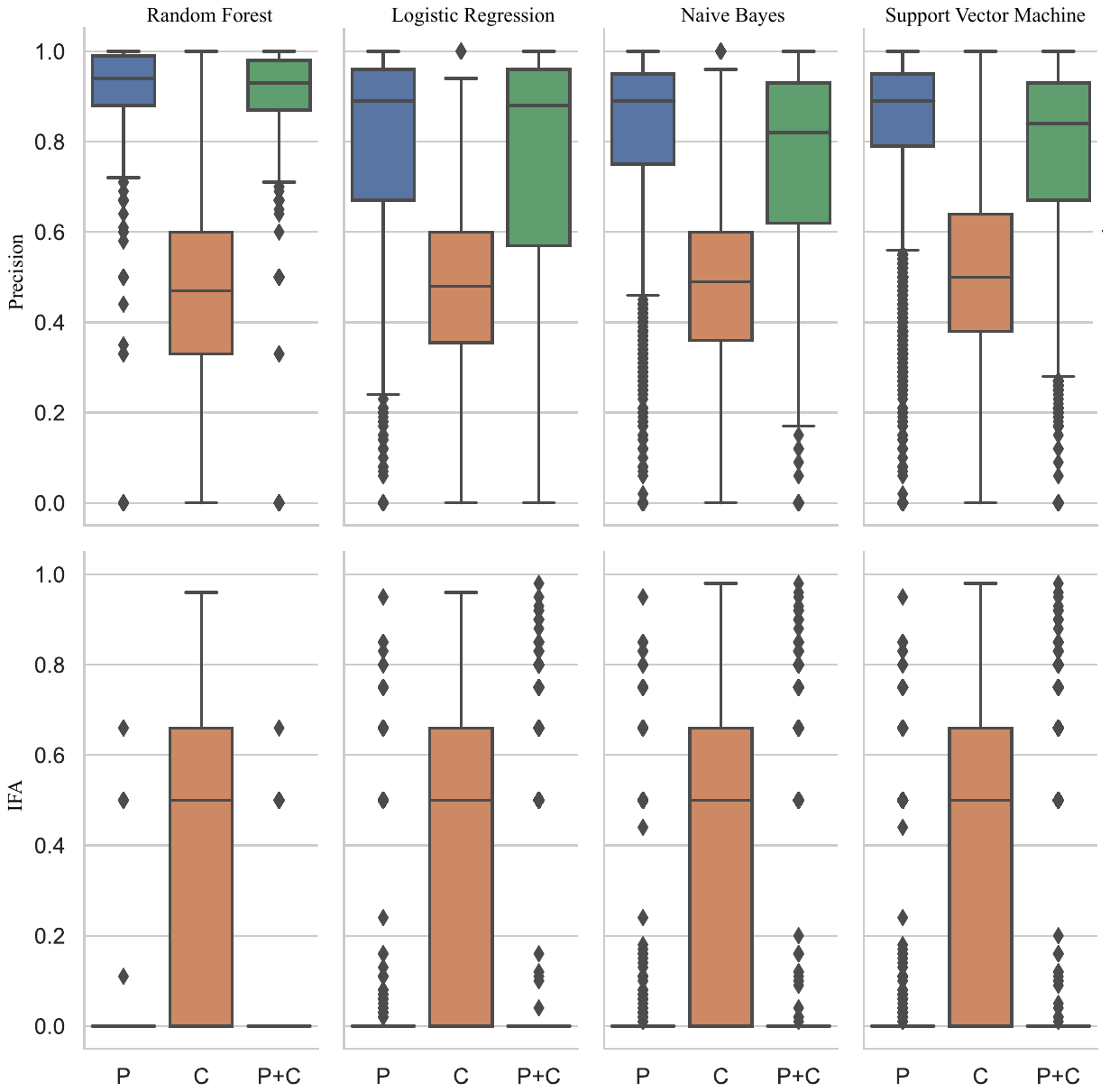}
    \caption{Cross-validation IFA and precision results  for  Process(P), Product(C), and Combined (P+C) metrics.
    Same format as \fig{learner_performance_1}.} 
    \label{fig:learner_performance_3}
\end{figure}

\begin{figure}[!b]
\centering
\begin{subfigure}[t]{\textwidth}
\centering
    \includegraphics[width=.8\linewidth]{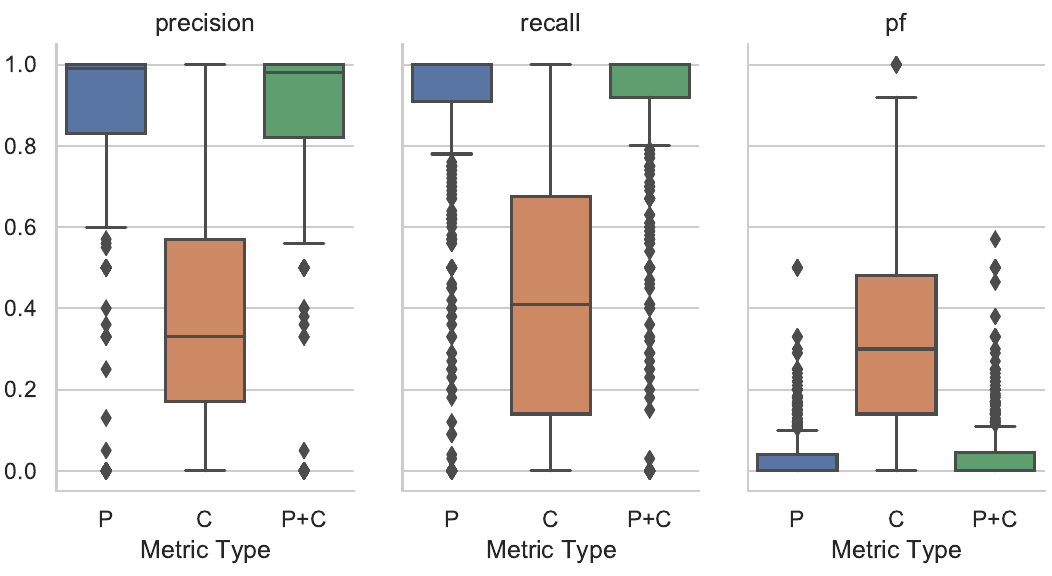}
\end{subfigure}
\begin{subfigure}[t]{\textwidth}
\centering
    \includegraphics[width=.8\linewidth]{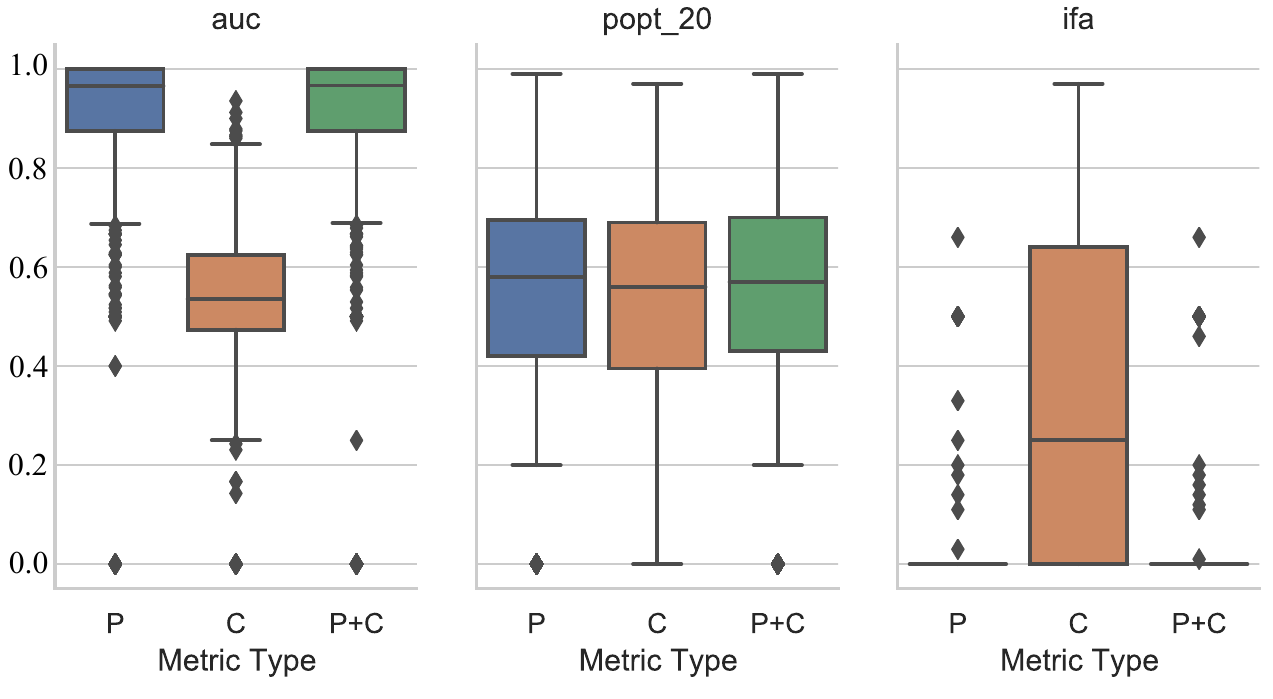}
\end{subfigure}

    \caption{Release based results for Random Forests. Here the training data was till t-3 th release and the rest was test release.} 
    \label{fig:release}
\end{figure}

To answer this question, we use Figure~\ref{fig:learner_performance_1}, Figure~\ref{fig:learner_performance_2}, Figure~\ref{fig:learner_performance_3}, and Figure~\ref{fig:release} to compares Recall, Pf, AUC, Popt20, Precision, and IFA across four different learners using process, product, and combined metrics. In those figures, the  metrics are marked as P (process metrics), C (product metrics), and combined (P+C). Figure~\ref{fig:learner_performance_1}, Figure~\ref{fig:learner_performance_2}, and Figure~\ref{fig:learner_performance_3} represents the cross-validation results, while Figure~\ref{fig:release} represent the release-based results.

For this research question, the  key thing to watch in these figures is the vertical colored box plots. The box plots were generated using results from all 700 Github projects, where each data point for a project is the (a)~median result from 5-fold cross-validation repeated 5 times for Figure~\ref{fig:learner_performance_1}, Figure~\ref{fig:learner_performance_2}, Figure~\ref{fig:learner_performance_3}, and (b)~median result from 3 release for Figure~\ref{fig:release}. These horizontal lines running across their middle show the median performance of a learner across 700 Github projects. As we said above in  section~\ref{sec:eval}, the best learners are those that {\em maximize} recall, precision, AUC, Popt20 while {\em minimizing} IFA and false alarms.
 
Reading the median line in the box plots, we say that compared to the Rahman et al. analytics in-the-small study, this analytics in-the-large study says some things are the same and some things are different. Like Rahman et al., these results show  clear evidence of the superiority of process metrics since, except for Popt20 (no significant difference across process, product, and process+product metrics) across all learners, the median process results from process metrics are clearly always better. That is to say, returning to our introduction, this study strongly endorses the Hersleb hypothesis that how we build software is a major determiner of how many bugs we inject into that software.

As to where we differ from the prior analytics in-the-small study, Random Forest with process metrics is statistically significantly better (achieving different statistical rank in Scott-Knott test) than any learner in all performance measure, other than Popt20 and IFA. In the case of Popt20 and IFA, all learners achieve the same statistical ranking from the Scott-Knott test. With these results we need to keep in mind, the Logistic Regression, Naive Bayes, and Support Vector Machine were tuned using hyper-parameter optimization, while the result for Random Forest was using default parameters. Thus the hyper-parameter tuned Logistic Regression and Support Vector Machine models were much costlier to build (256 hours for hyper-parameter tuned Support Vector Machine for vs 10 hours for default Random Forest). So, unlike the Rahman et al. analytics in-the-small study,  we would argue that it is very important which learner is used to for analytics in-the-large. Certain learning in widespread use such as Naive Bayes, Logistic Regression, and Support Vector Machines may not be the best choice for reasoning from hundreds of software projects. Rather, we would recommend the use of Random Forests.

We also performed a small experiment to see if certain metrics only capture certain defects as part of this study.   We analyzed the defects that are only captured by process metrics vs the defects that are only captured by product metrics. Looking into our results,
we see that:
\bi
    \item Process metrics capture nearly all the defects; evidence: see the very high recall scores for Random Forest process metrics in Figure~\ref{fig:learner_performance_1}.
    \item As to product metrics, they tended to miss many defects; observe how, for all learners in  Figure~\ref{fig:learner_performance_1}, the product metrics recall are much lower than than the process metrics. For example. in the case of Random Forests, we found that  the product metrics missed 48\% of the defects found by process metrics,
\ei

On the other hand, there are indeed a  small number of defects captured by product metrics and not process metrics. But this case is definitely in the minority (less than 1\% in all our studies). Hence we say that  process metrics are superior at finding nearly all types of defects in a software system, while product metrics are not able to do that.

Before going on, we comment on certain other aspects of these results:
\bi
    \item We see   no evidence of  any added value of combining process and product metrics. If we compare the (P+C) results to the (P) results, there is no case in Figure~\ref{fig:learner_performance_1}, Figure~\ref{fig:learner_performance_2},  and Figure~\ref{fig:learner_performance_3} where process + product (P+C) metrics do better than just using process (P) metrics.
    \item Similar to Kamei et al. in the case of effort-aware  evaluation criteria process metrics are superior to product metrics, as can be seen in Figure~\ref{fig:learner_performance_3}. Note in that figure, many of our learners using process metrics have near-zero IFA scores. This is to say that, using process metrics, programmers will {\em not} be bombarded with numerous false alarms. But unlike Kamei et al., we do not see any significant benefit when accessing the performance in regards to the Popt20, which is another effort-aware  evaluation criteria used by Kamei et al. and this study.  
    \item \fig{release} shows the Random Forest results using release-based test sets.  As stated in section~\ref{sec:eval} above, there is very little difference in the results between release-based test generation and the cross-validation method o Figure~\ref{fig:learner_performance_1} and Figure~\ref{fig:learner_performance_2}, and Figure~\ref{fig:learner_performance_3}. Specifically, in both our cross-val and release-based results, (a)~process metrics do best; (b)~there is no observed benefit in adding in product metrics and, when using process metrics then random forests have (c)~very high precision and recall and AUC, (d)~low false alarms; and (e)~very low IFA.
\ei

\begin{RQbox}
    \textbf{RQ 2:} Measured in terms of predication variability, do methods that works well in-the-small, also work at at-scale?
\end{RQbox}
To answer this research question, we assess our learners not by their median performance but by their variability.

Rahman et al. commented that many different learners might be used for defect prediction since, for the most part, they often give the same results. While that certainly holds for their analytics in-the-small case study, the situation is very different when reasoning at-scale about 700 projects. Looking at the process metrics results for Figure~\ref{fig:learner_performance_1} and Figure~\ref{fig:learner_performance_2} and Figure~\ref{fig:learner_performance_3}, we see that -

\begin{enumerate}
    \item The performance for Random Forests is statistically significantly better in case all performance measures, other than Popt20 and IFA.
    \item The  box plots for Random Forests are much smaller than for  other learners in the case of precision, recall, and AUC. That is, the variance in the predictive performance is much smaller for Random Forest than for anything else in this study. 
    \item These results for Random Forests are without hyper-parameter optimization, while other learners are optimized with hyper-parameter optimization. This makes the model building for Random Forest orders of magnitude faster.
\end{enumerate}

The size of both these effects is quite marked. Random Forest is usually better (median) than Logistic Regression. As to the variance, the Random Forest variance is  {\em smaller} than the other learners.

Why is Random Forest doing so well?  We conjecture that when reasoning about 700 projects that there are many spurious effects. Since Random Forests make their conclusions by reasoning across multiple models, this kind of learner can avoid being confused. Hence, we recommend ensemble methods like Random Forest for analytics in-the-large. 

\begin{RQbox}
    \textbf{RQ 3:} Measured in terms of granularity, do same granularity that works well in-the-small, also work at at-scale?
\end{RQbox}

In this research question, we try to evaluate if the granularity of the metrics matters when predicting for defects when measuring at scale. This is one of the research questions asked in study by Kamei et al.. Here we try to measure if package-level prediction better identifies defective packages than file-level prediction. There are multiple strategies for creating package-level metrics such as lifting file-level metrics to package-level, collecting metrics designed for package-level, and lifting file-level prediction results for package-level as explored by Kamei et al. in their study. We explore the first strategy that is to lift the file-level metrics to package-level. We select this strategy as Kamei et al. in their paper has shown the metrics designed for package-level does not produce good results and both file and result lift ups have similar performance and have been explored by many other researchers. To build a defect predictor using package-level data, we use the process metrics collected for our tasks. For each commit/release, if there are multiple files from the same package, we aggregate them to their package-level by taking the median values. 

\begin{figure*}[!t]
\begin{subfigure}[t]{\textwidth}
\centering
    \includegraphics[width=.8\linewidth]{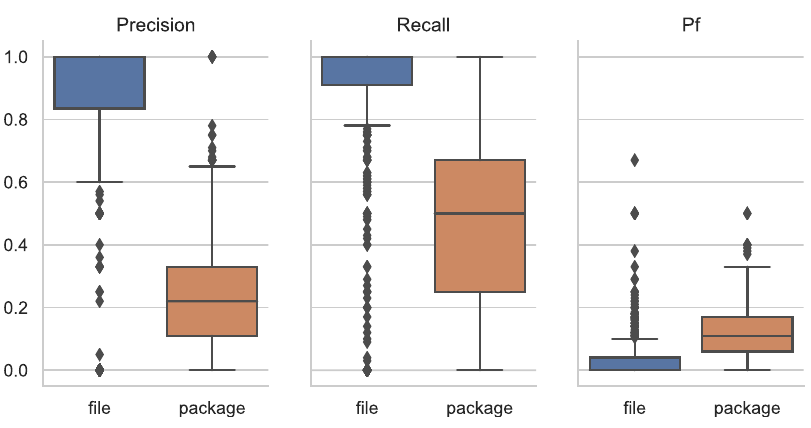}
\end{subfigure}
\begin{subfigure}[t]{\textwidth}
\centering
    \includegraphics[width=.8\linewidth]{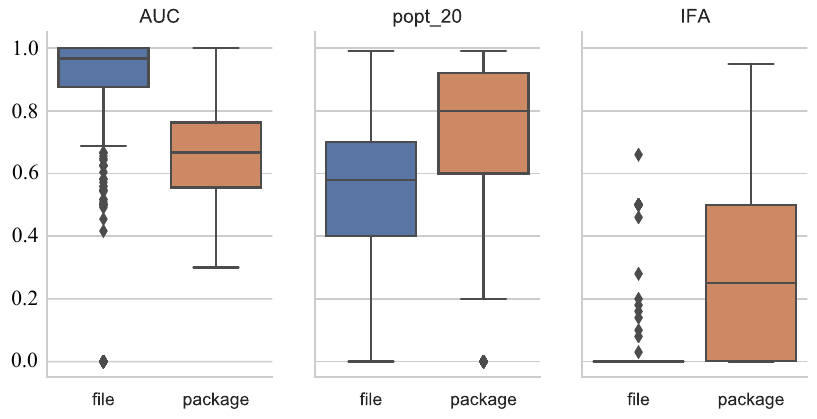}
\end{subfigure}
    \caption{File vs package-level prediction for models built using file-level process data and package-level process data.} 
    \label{fig:RQ3}
\end{figure*}

Figure~\ref{fig:RQ3} shows the difference in performance between file-level prediction results and package-level prediction results. It is evident from the results, that file-level prediction shows statistically significant improvement than package-level prediction, with an exception in the case of Popt20. This result agrees with Kamei et al., and we conclude  that the granularity of the metrics set does matter and file-level level prediction has superior performance than package-level prediction.

\begin{figure}[]
\begin{subfigure}[t]{\textwidth}
\centering
    \includegraphics[width=0.8\linewidth]{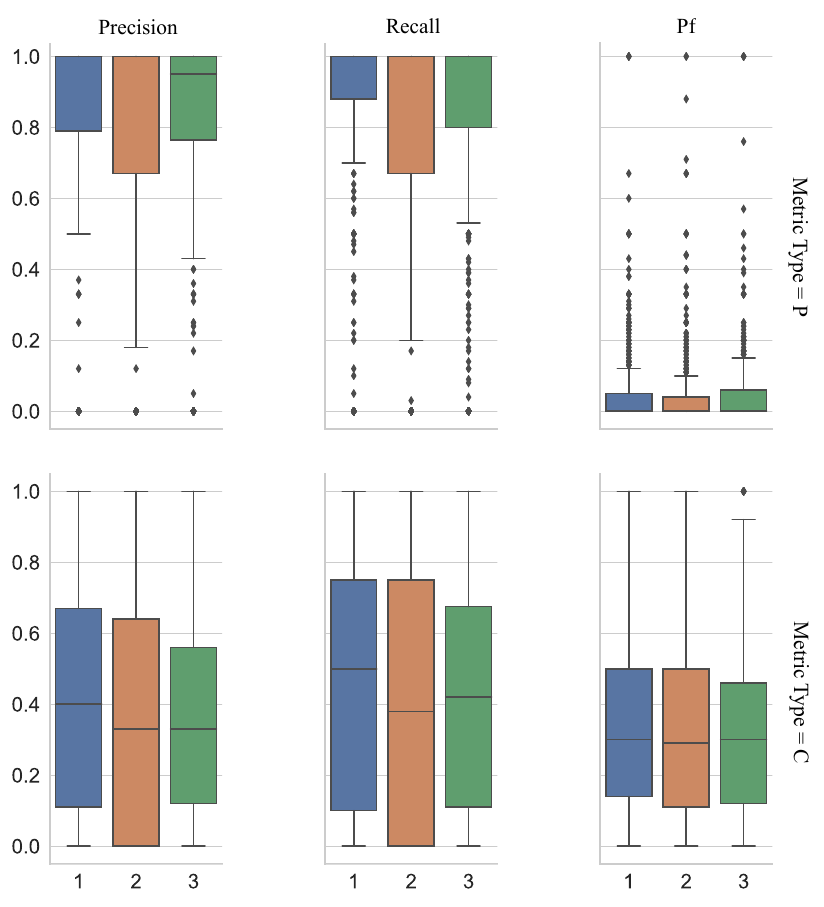}
\end{subfigure}
\begin{subfigure}[t]{\textwidth}
\centering
    \includegraphics[width=0.8\linewidth]{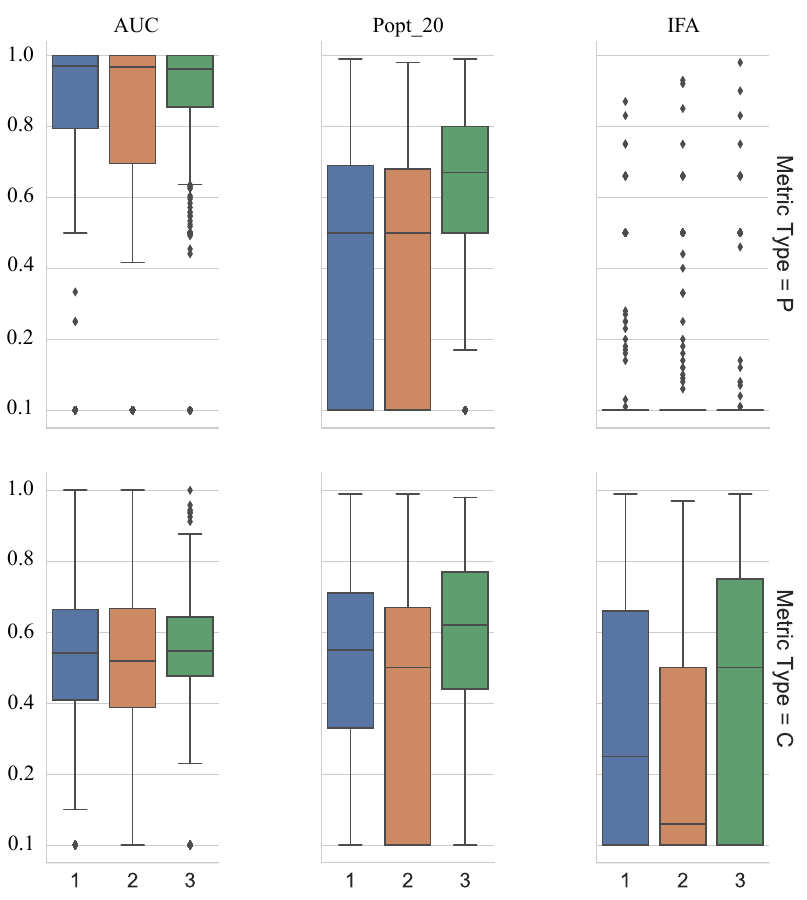}
\end{subfigure}
    \caption{Stability of the models across the last 3 releases built using process (P) and product (C) metrics. Each plot shows one of the six performance criteria used in this study for the last 3 releases. The first row shows the results for the process metrics denoted as {\em Metric Type = P} and the second row shows the results for product metrics denoted as {\em Metric Type = C}. } 
    \label{fig:stability}
\end{figure}

\begin{RQbox}
\textbf{RQ 4:} Measured in terms of stability, are process metrics more/less stable than code metrics, when measured at at-scale?    
\end{RQbox}

To answer this research question, we first tag each commit into a release by using the release information from Github. Using this release information, we divide the data into train and test data using the last 3 releases as test releases one by one and other older releases as training data. If a model build using either process and product data significantly differ across last 3 releases, that would imply the model built using that set of metrics will need to be rebuilt for each subsequent release, this in-tern will create instability. To verify the stability of the models built using metrics, we build the models using the training data and then check each of the 3 subsequent releases in term of the evaluation criteria used in this study. We compare both process and product metrics across all 6 criteria mentioned in Section~\ref{sec:eval}. 

Figure~\ref{fig:stability} shows the performance of the models. The first row of the figure represents the process metrics, while the second row represents the product metrics. Each column represents the evaluation criteria that we are measuring and inside each plot, each box plot represents one of the last 3 releases. We applied Scott-Knott statistical test on the results to check for each evaluation criteria if any of the releases are statistically significantly different than the others. The results show no significant difference between 3 releases in all evaluation criteria (all releases for each evaluation criteria in each metric type) except Popt20.  Popt20 is an effort-aware  criterion as explained in Section~\ref{sec:eval}, and we see in both process-based and product-based models the  Popt20 does significantly better in the third release. Which may be because third release have more smaller predicted defective files than two releases. If that is the case, based on how Popt20 is calculated it can explain the increase in Popt20 score. That being said, the result shows none of the models build using process and product metrics degrades over time, thus reducing the instability of the models. We can also say, as over time, the performance does not degrade and we have already seen in terms of performance process metrics performs much better than product metrics, it is wiser to use process metrics in predicting defects.

% \begin{figure}[]
% \begin{subfigure}[t]{\textwidth}
% \centering
%     \includegraphics[width=0.8\linewidth]{RF_stability_1.pdf}
% \end{subfigure}
% \begin{subfigure}[t]{\textwidth}
% \centering
%     \includegraphics[width=0.8\linewidth]{RF_stability_2.pdf}
% \end{subfigure}
%     \caption{Stability of the models across the last 3 releases built using process (P) and product (C) metrics. Each plot shows one of the six performance criteria used in this study for the last 3 releases. The first row shows the results for the process metric denoted as {\em Metric Type = P} and the second row shows the results for product metric denoted as {\em Metric Type = C}. } 
%     \label{fig:stability}
% \end{figure}

\begin{RQbox}
\textbf{RQ 5:} Measured in terms of stasis, Are process metrics more/less static than code metrics, when measured at at-scale?    
\end{RQbox}

In this research question, we try to find the reason behind the difference in performance in models built using process and product data. Most models try to learn how to differentiate between two classes by learning the pattern in the training data and tries to identify similar patterns in the test data to predict for defects.  Throughout the life cycle of a project, different parts of the project are updated and changed as part of regular enhancements. This results in introduction of bugs and thus bug fixes for those defective changes.  The metrics that we use to create the defect prediction models should be able to reflect those changes, so the model is able to identify the difference between defective and non-defective changes. This means if either process or product metrics can capture such differences, then the metric values for a file between release $ R $ and $R + 1$ would not be highly correlated, and models built with such metrics will be able to better differentiating  defective and non-defective change.

\begin{figure}[!t]
\centering
    \includegraphics[width=0.8\linewidth]{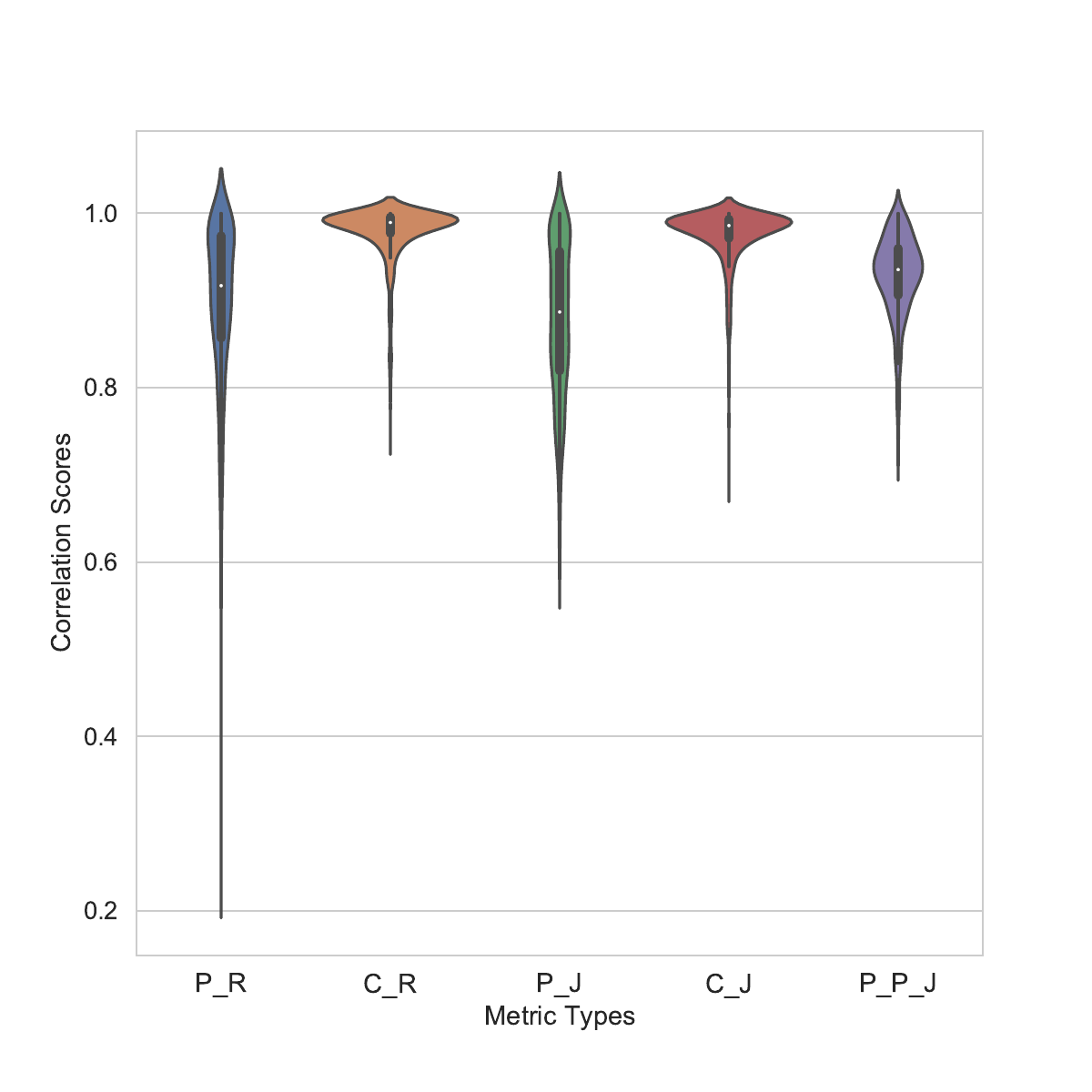}
    \caption{The plot represents the Spearman  correlation of every file between two consecutive checkpoints. Here x-axis label P\_R and C\_R represents the process and product metrics when the correlation was calculated in release level. While the P\_J, C\_J, and P\_P\_J represent the process, product, and package-level process metrics when calculated in JIT-based  setting.}
    \label{fig:corr}
\end{figure}

To measure the stasis of the metrics, we used Spearman correlation for every file between two consecutive releases (to check releases-based prediction) and two consecutive commits where the file was changed (to check for JIT-based  predictions). Here the metrics for each file for a release are calculated from the last time the file was changed before the release. Thus for comparing between release $ R $ and $R + 1$ for a file, we select the commit the file was changed last both for release $ R $ and $R + 1$ and compute the Spearman correlation between them. Figure~\ref{fig:corr} shows the Spearman correlation values for  every file between two consecutive releases/commits for all the projects explored as a violin plot for each type of metric. A wide and short violin plot represents the majority of the value concentrated near a certain value. In contrast, a thin and long violin plot represents values being in a different range. Figure~\ref{fig:corr} shows the correlation scores for process and product metrics in both release-based and JIT-based  settings. The process and product metrics in release-based settings are denoted by P\_R and C\_R respectively, while in JIT-based  setting they are denoted by P\_J and C\_J respectively. In the figure, the P\_P\_J represents the package-level process metrics in JIT-based  setting. We can see from figure~\ref{fig:corr}, the product metrics form a wide and short violin plot and are very highly correlated. While the process metrics form a thin and long violin plot ranging between 0.2 to 1 for release-based setting and 0.5 to 1 for JIT-based  setting. If we compare the correlations between release-based and JIT-based  metric sets, we see the correlation value for process metrics increases in JIT-based  metric sets. The reason behind this increase in correlation value can be explained as in JIT-based  metrics, we compare between commits. Here the amount of the change in file is less than the change when measured between two releases (here each release contains multiple commits).  Similarly, when the process metrics has been lifted from file-level to package-level, the correlation increases.

So why process metrics outperform product metrics? We think the stasis property of the metric set is one of the main reasons as product metrics seems to be more static, thus changing very little with time and between defective files and non-defective files. When models are created with such static metric sets, it is hard for the model to learn a pattern and differentiate between defective and non-defective changes. While process metrics change over time and much less correlated between changes, thus making them a potentially better metric for creating defect prediction models.

\begin{figure}[!b]
\centering
    \includegraphics[width=0.6\linewidth]{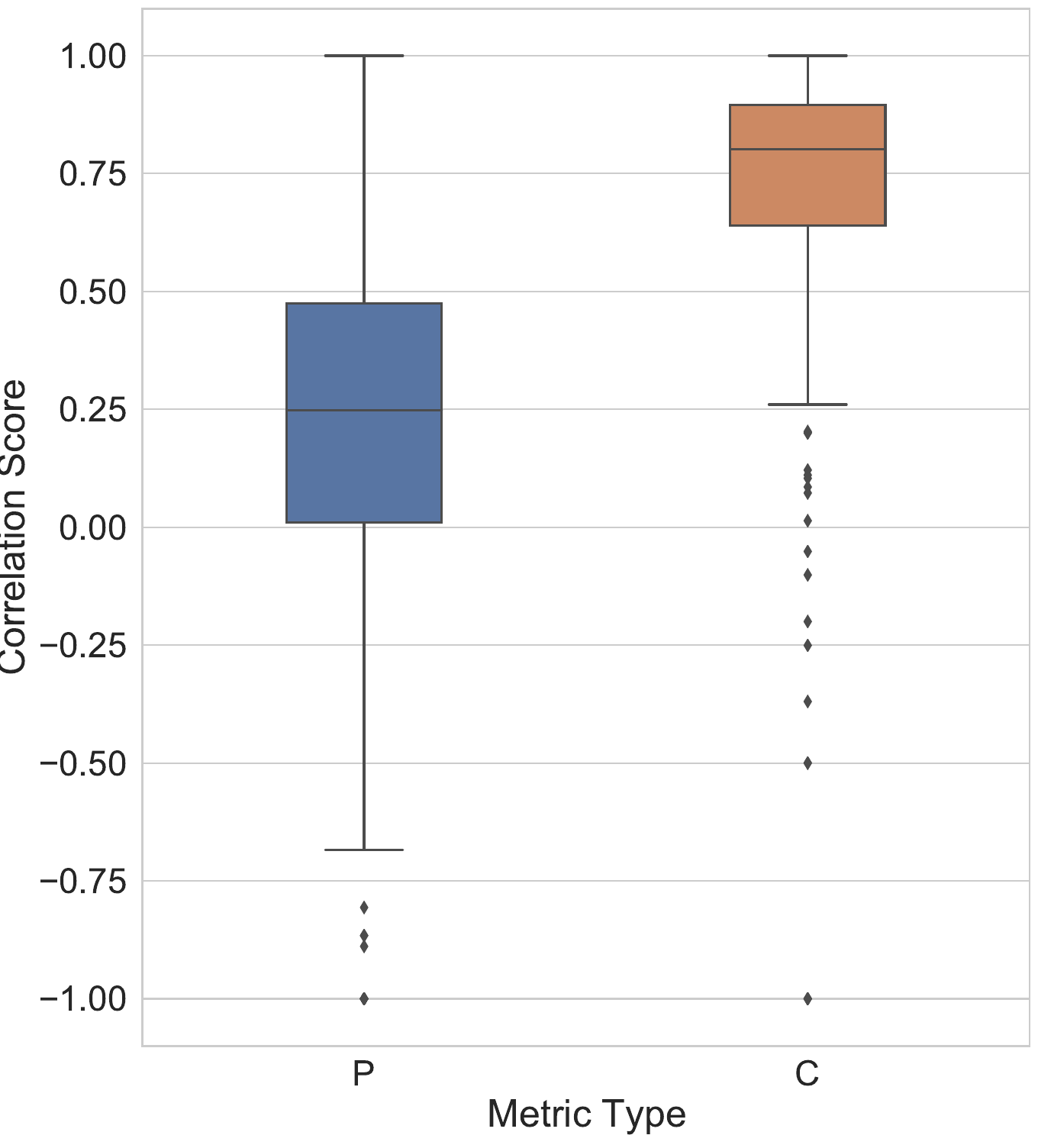}
    \caption{The plot represents the Spearman  correlation between probabilities of defect-proneness across all pairs of training-test releases. }
    \label{fig:rq6}
\end{figure}

\begin{RQbox}
\textbf{RQ 6:}  Measured in terms of stagnation, Do models built from different sets of metrics stagnate across releases, when measured at at-scale?
\end{RQbox}

In this research question, we try to measure the stagnation property of the models built using the process and product metrics. As suggested by Rahman et al., we use Spearman rank correlation between the learned probability from the training set  and predicted probability from the test set to calculate the correlation between these two. To learn the learned probability and predicted probability, we use the defect-proneness from the learner (Random Forest in this research question) across all pairs of training-test releases. For each pair of training-test releases, if a file has been committed multiple times during a release, we consider the file instance that was changed last. Here a high correlation between the learned and predicted probability, which will indicate the models are probably learning to predict the same set of files defective. It is finding the same probabilities in the test set as training set and thus, it is not able to properly differentiate between defective and non-defective files. Figure~\ref{fig:rq6} shows a box plot of Spearman rank correlation between the learned and predicted probability for models built using process and product metrics on 700 projects used as part of this study. We can see that, a model built using product data has significantly higher correlation than a model built using process data. Although this value is slightly lower, both in the case of process and product metrics  than what Rahman et al. reported in their project, the results signify the models built with product metrics are significantly more stagnant than the models built using process metrics.

% \begin{figure}[!t]
% \centering
%     \includegraphics[width=0.6\linewidth]{RQ6.pdf}
%     \caption{The plot represents the Spearman  correlation between probabilities of defect-proneness across all pairs of training-test releases. }
%     \label{fig:rq6}
% \end{figure}

\begin{figure}[!b]
\centering
    \includegraphics[width=0.8\textwidth]{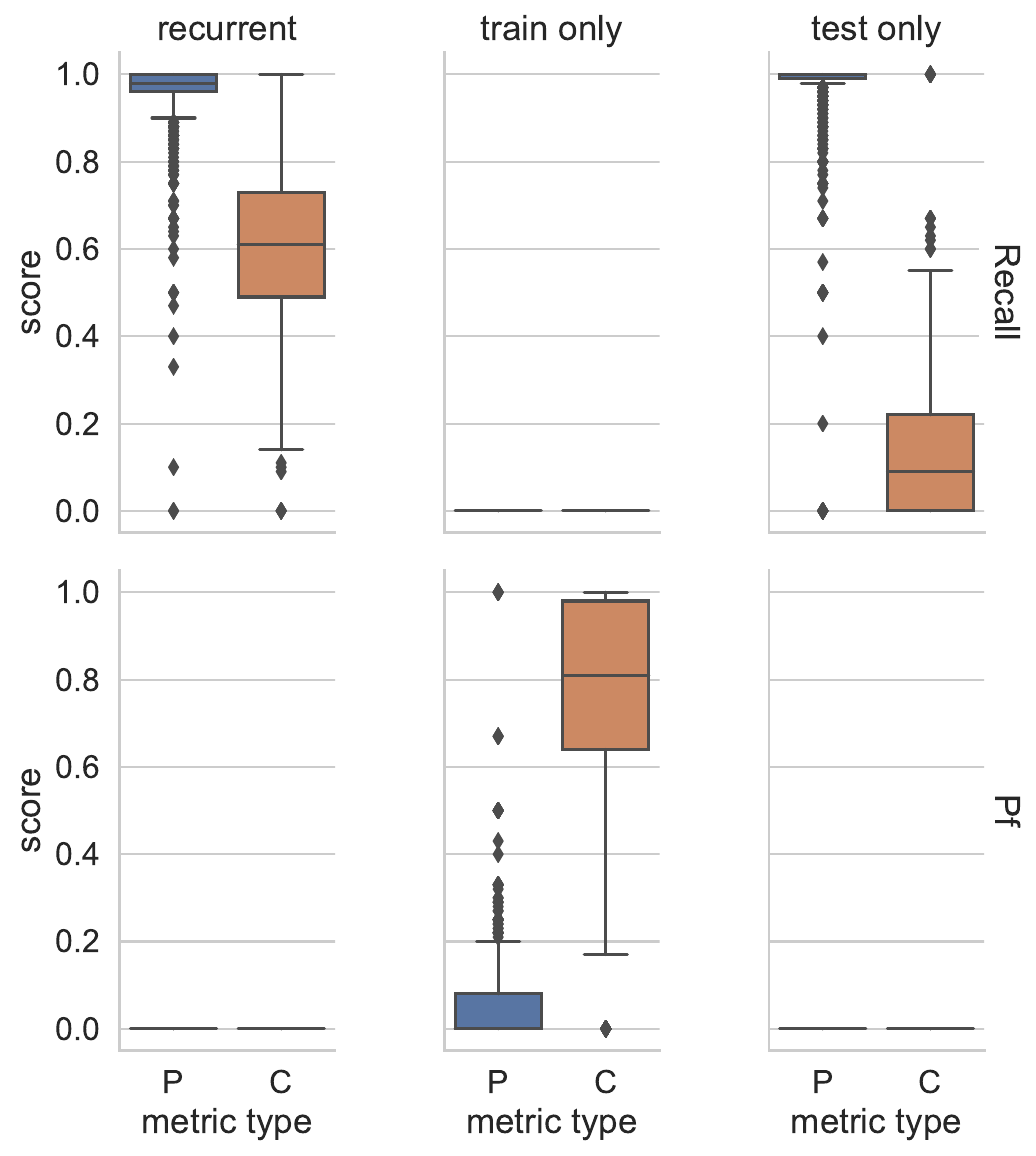}
    \caption{Performance of the models build using process and product metrics on recurrent, train only and test only test sets.}
    \label{fig:RQ7}
    % \vspace{-180pt}
\end{figure}

\begin{RQbox}
\textbf{RQ 7:} Do stagnant models (based on stagnant metrics) tend to predict recurringly defective entities?  
\end{RQbox}

Here we try to verify the stagnation property of the metrics as seen in the previous research question. If a model is stagnant, it will predict the same file as defective regardless of whether the file actually contains defects or not. To evaluate whether or not model built on process and product data is predicting the same files as defective, we follow the same approach suggested by Rahman et al. For each training test pairs, if there are multiple instances of the same file in a release, we select the last instance when it was changed for both training and test data. We then divide the test data into 3 parts (a) part 1 only contains files that are defective in both training and test set, we call this recurrent  set (b) part 2 consists of files that are defective in the training set but not in the test set, this is train only set and finally (c) part 3 only contains files that are defective in the test and not in the training set, we call this test only set. A model, if stagnant, will have a high recall for recurrent set, high pf for train only set, low recall for test only set and that will show the model is actually predicting the same set of files as defective and not able to identify new defective files. Figure~\ref{fig:RQ7} shows the recall and pf of the models build using process and product metrics on all 3 types of test sets. We can see from the figure that models built using either process or product metrics can identify recurrently defective files in case of recurrent set. However, we can see a significant difference between process and product metrics, where process metrics is doing much better in recognizing recurrently defective files. In case of train only test set, we can see very high pf (median value $\approx 0.8$) for model build using product data, while the model built using process data has a low pf (median value $\approx 0.0$). This is a clear indication that model built using product metrics is stagnant and identifies the same set of files as defective regardless of whether they are actually defective or not. While the test only set shows a very low recall for model built using product data, while high recall for model built using process data. This indicates model built using product data is unable to identify new defects. Thus this result bolsters the claim that process metrics are better at identifying defects than product metrics.

% \begin{figure}
% \centering
%     \includegraphics[width=0.8\textwidth]{RF_RQ6.pdf}
%     \caption{Performance of the models build using process and product metrics on recurrent, train only and test only test sets.}
%     \label{fig:RQ7}
%     % \vspace{-180pt}
% \end{figure}

\begin{RQbox}
\textbf{RQ 8}:
 Measured in terms of metric importance, are metrics that seem important in-the-small, also important when reasoning in-the-large?
\end{RQbox}

\begin{figure}[]
\centering
    \includegraphics[width=\linewidth]{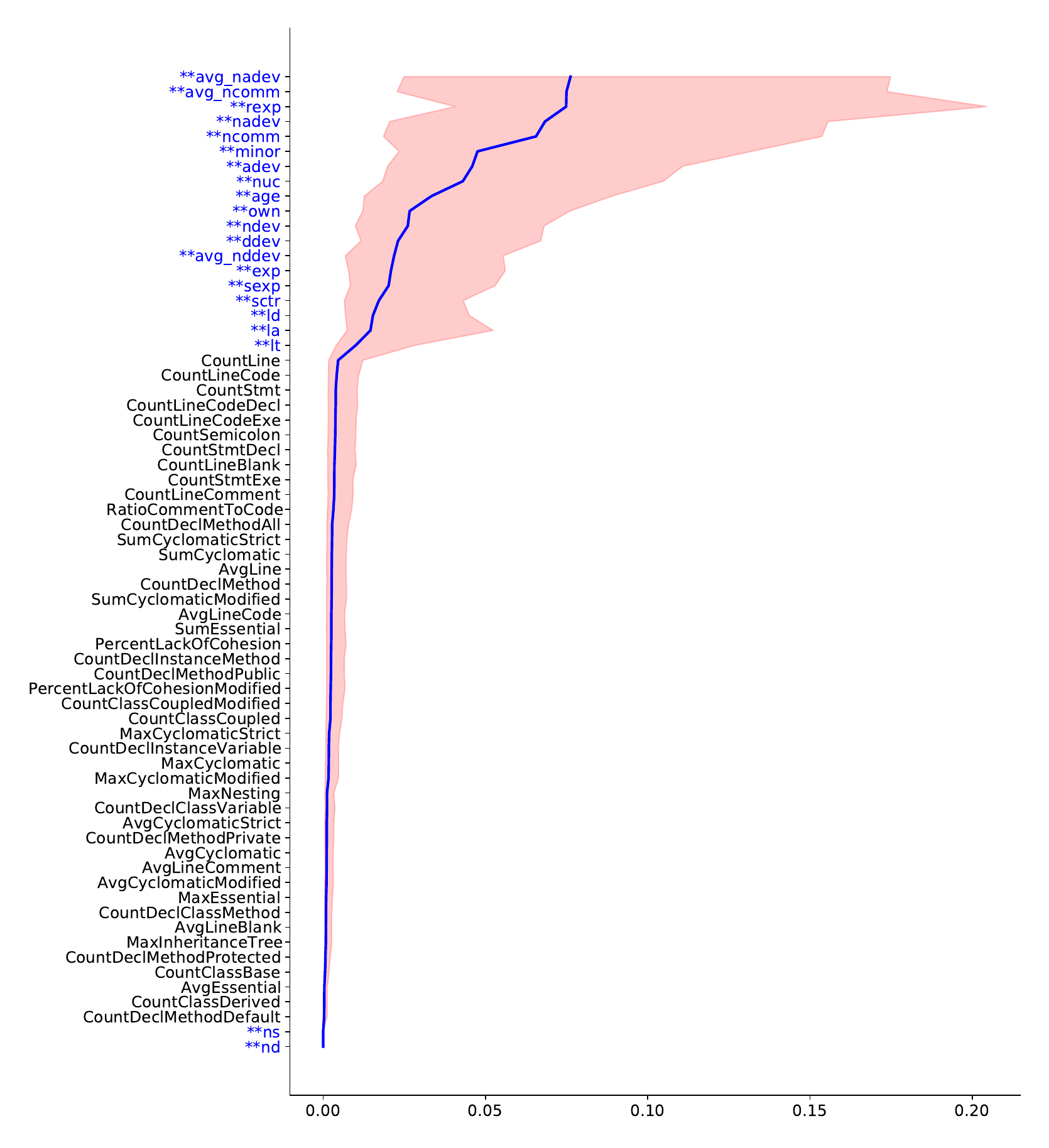}
    \caption{Metric importance of process+product combined metrics based on Random Forest.
    Process metrics are marked with \textcolor{blue}{two blue asterisks**}. Blue denotes the median
    importance in 700 projects while the pink region shows the (75-25)th percentile.} 
    \label{fig:process+product_feature}
\end{figure}

To answer this question, we test   if what is learned   from studying {\em some} projects is the same as what might be learned from studying {\em all} 700 projects. That is,  we compare the rankings given to process metrics using all the projects (analytics in-the-large) to the rankings that might have been learned from analytics in-the-small projects looking at 5 projects (where those projects were selected at random).

Figure~\ref{fig:process+product_feature} shows the metric importance of metrics in the combined (process + product) data set. This metric importance is generated according to what metrics are important while building and making predictions in Random Forest. The metric importance returned by Random Forest is calculated using a method implemented in Scikit-Learn. Specifically: how much each metric decreases the weighted impurity in a tree. This impurity reduction is then averaged across the forest and the metrics are ranked. In Figure~\ref{fig:process+product_feature} the metric importance increases from left to right. That is,   in terms of defect prediction, the most   important metric is the average number of developers in co-committed files (avg\_nadev) and   the least important metric is  the number of directories (nd).

In that figure, the process metrics are marked with \textcolor{blue}{two blue asterisks**}. Note that nearly all of them appear on the top. That is, in a result consistent with Rahman et al., process metrics are far more important than process metrics.  

\begin{figure}
\centering
    \includegraphics[width=0.8\linewidth]{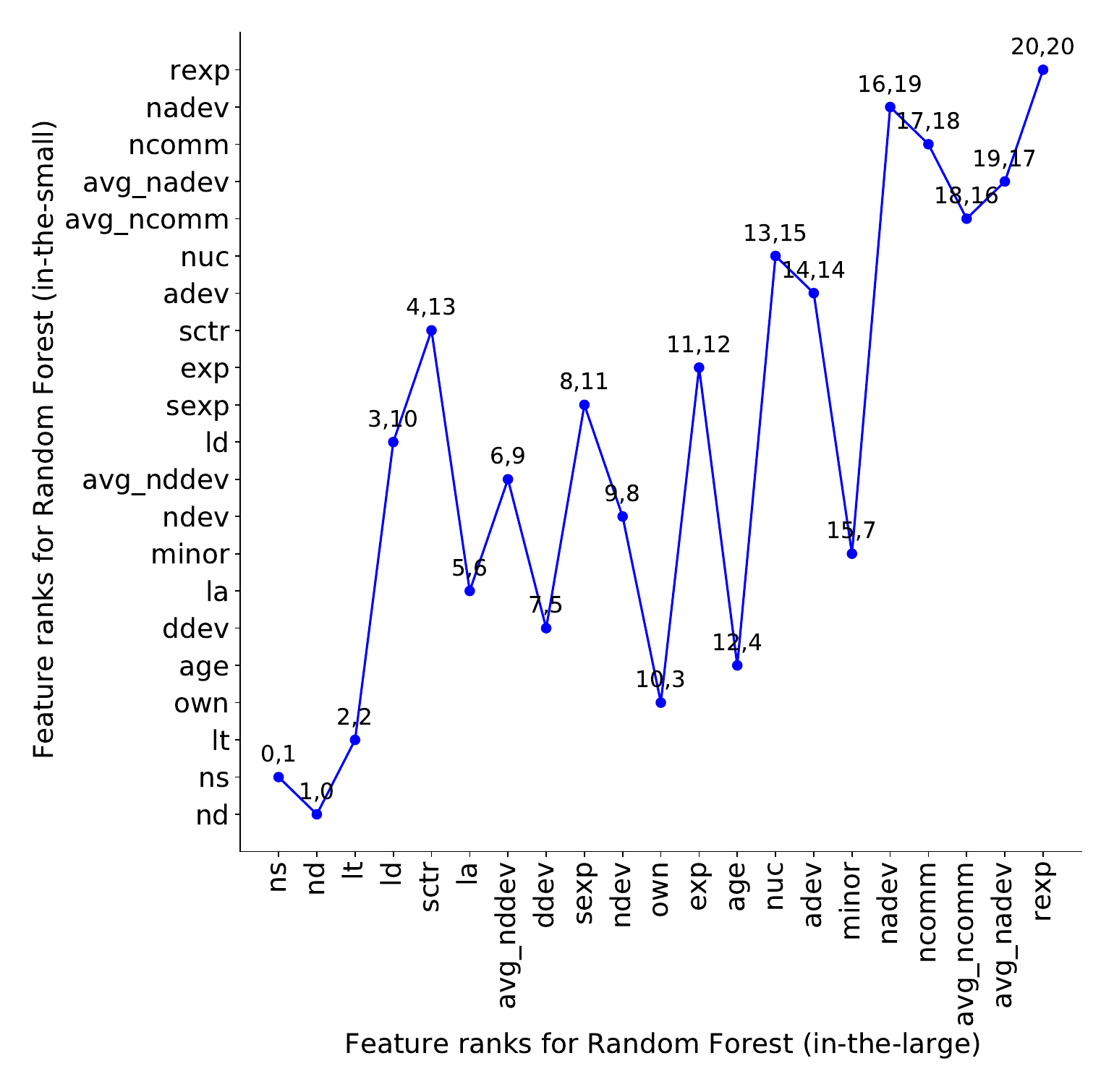}
    \caption{X-axis ranks metrics via analytics in-the-large (using Random Forests applied to 700 projects).
    Y-axis ranks process metrics using  analytics in-the-small (using Random Forests selected from random sample of 5 projects).} 
    \label{fig:rank_attr}
\end{figure}

Figure~\ref{fig:rank_attr} compares the process metrics rankings learned from analytics in-the-large (i.e., from 700 projects) versus a simulation of an in-the-small study that looks at five projects selected at random. In the figure, the X-axis ranks metrics via analytics in-the-large (using Random Forests applied to 700 projects), and Y-axis ranks process metrics using  analytics in-the-small (using Random Forests applied to randomly select 5 projects). For both x and Y-axis rankings, the metrics were sorted by the metric importance returned by the Random Forest Classifier. 
    
In an ideal scenario,   when the ranks are the same, this would appear in Figure~\ref{fig:rank_attr} as  a straight line at a 45-degree angle, running through the origin. To say the least, this {\em not} what is observed here. We would summarize Figure~\ref{fig:rank_attr} as follows: the importance given to metrics by a few analytics in-the-small studies is very   different from the importance learned via  analytics in-the-large.

\section{THREATS TO VALIDITY}
\label{sec:threats}

As with any large scale empirical study, biases can affect the final results. Therefore, any conclusions made from this work must be considered with the following issues in mind:

(a) \textit{Evaluation Bias}: 
In all research questions in this study, we have shown the performance of models built with process, product and, process+product metrics and compared them using statistical tests on their performance to conclude which is better and more generalizable predictor for defects. While those results are true, that conclusion is scoped by the evaluation metrics we used to write this paper. It is possible that using other measurements, there may be a difference in these different kinds of projects (e.g., G-score, harmonic mean of recall, and false-alarm reported in \cite{Tu20_emblem}). This is a matter that needs to be explored in future research.

(b) \textit{Construct Validity}: At various places in this report, we made engineering decisions about (e.g.,) choice of machine learning models,  selecting metric vectors for each project. While those decisions were made using advice from the literature, we acknowledge that other constructs might lead to different conclusions. 

(c) \textit{External Validity}: For this study, we have collected data from 700 Github Java projects. The product metrics collected for each project were done using a commercialized tool called ``Understand'' and the process metrics were collected using our own code on top of Commit\_Guru repository. There is a possibility that calculation of metrics or labeling of defective vs non-defective using other tools or methods may result in different outcomes.  That said, the ``Understand'' is a commercialized tool with detailed documentation about the metrics calculations. We have shared our scripts and processes to convert the metrics to a usable format and has described the approach to label defects.  

(d) \textit{Sampling Bias}: Our conclusions are based on the 700 projects collected from Github. It is possible that different initial projects would have lead to different conclusions. That said, this sample is very large, so we have some confidence that this sample represents an interesting range of projects.  

(e) \textit{Selection Bias:} Our comparison between  process, product and, process+product metrics are based on metrics used in prior work (Rahman et al.~\cite{rahman2013and}   Kamei et al.~\cite{kamei2010revisiting}). It is certainly true  that  other metrics might be more important than those explored here. For future work, we strongly recommend exploring a wider range of metrics; e.g., such as those suggested by other researchers~\cite{radjenovic2013software, pascarella2020performance, li2018progress}.

\section{CONCLUSION AND DISCUSSION}
\label{sec:conclusion}

Much prior work in software analytics has focused on in-the-small studies that used a few dozen projects or less.  Here we checked what happens when we  take specific conclusions generated from analytics in-the-small, then review those conclusions using analytics in-the-large. While some conclusions remain the same (e.g., process metrics generate better predictors than product metrics for defects), other conclusions change (e.g., learning methods like logistic regression that work well in-the-small perform comparatively much worse when applied in-the-large).

We find here that
issues that may not seem critical in-the-small become significant problems in-the-large. For example:
\bi
\item
Recalling \fig{rank_attr}, we can say that  what seems to be an important metric, in-the-small, can prove to be very unimportant when we start reasoning in-the-large. 
\item
Further, when reasoning in-the-large,  variability in predictions becomes a concern.
\ei
Thus when researchers or industry practitioners 
attempt to:
\bi
\item
Generate guidelines or best practices to either train new researchers or developers;
\item
Create tools for quality measurements, guide developers to follow best practices or helping developers or researchers in other ways; 
\item
Study data to find general defect-related trends/properties of open-source projects;
\ei
then it is better to use findings from in-the-large analysis. The reason being, if the lessons learned change from project to project, it will be very hard to generate guidelines or create tools that are stable enough for an organization.
This is an issue since:
\bi
\item
If the guidelines or tools are not stable, then developers or researchers will lose trust in those tools.
\item
Also, when trying to find general trends in software projects, trends found from an in-the-small study might change when the selected projects are changed and thus, those will not be general trends but project specific trends. 
\item
We found that  certain systems issues seem unimportant in-the-small. However, when scaling up to in-the-large,  it becomes a critical issue that  product metrics are an order of magnitude to harder to manage. We listed one case study above where the systems requirements needed for product metrics meant that, very nearly, we almost did not deliver scientific research in a timely manner.
\ei
\noindent
Based on this experience, we say:
\bi
    \item Industrial practitioners should make use of in-the-large findings or re-validate in-the-small findings with in-the-large analysis before applying them to organizational level either to create guidelines or to make tools.
    \item Analysts performing analytics in-the-large should  use process metrics and ensemble methods like random forests since they  can better handle the kind of large scale spurious singles seen while reasoning effectively over hundreds of projects. 
    \item  SE researchers must now:
        \bi
            % \item Report in-the-large analysis when reporting new findings or compare their findings with previous in-the-large analysis;
            \item Revisit many of the conclusions previously obtained via analytics in-the-small to find if those findings still hold true for in-the-large analysis.
            \item Perform in-the-large analysis when trying to find general trends in software projects in their research.
        \ei
\ei

More generally, what is this work saying about the  notions/need/benefits  of quantitative  versus qualitative    in  defect-related  research  in-the-large? Quantitative studies can scale to a very large number of projects (as shown by this study), while qualitative studies can find specific, nuanced features that are specific to that small set of projects (evidence, see Figure~\ref{fig:rank_attr}). However, it would be wrong to use this study to say (e.g.,) ``stop qualitative studies'' since, in our experience, more can   be achieved by combining the two approaches (than just mono-focusing on just qualitative or quantitative).

For example, previously, with Chen and Stolee et al.~\cite{csm19}, we have  argued for a marriage of qualitative and quantitative methods to effectively reduce the effort associated with the partial replication and enhancement of qualitative studies. In the case study of that paper~\cite{csm19}, a qualitative study explored factors influencing the fate of GitHub pull requests using extensive qualitative analysis of 20 pull requests. Guided by their findings, we mapped some of their qualitative insights onto quantitative questions. To determine how well their findings generalize, we collected much more data (ten times as many additional pull requests from  hundreds of GitHub projects). This combined approach resulted in a new predictor for whether code would be merged. That predictor was far more accurate than one built from the study's qualitative factors (F1=90 vs 68\%),  illustrating the value of a mixed-methods approach and replication to improve prior results. We conjecture that that case study is representative of an underlying methodology for scaling and extending primary qualitative studies that require expert opinions. 

Hence, we argue that one future direction for this research could be to  encourage more studies that replicate parts of primary qualitative studies  using quantitative methods (since these scale to a large number of projects).
Further, we should not stop there. The insights gained from this combined qualitative/quantitative approach could be used to design insightful subsequent studies.

\section{Acknowledgments}
\label{sec:ack}
 This work was partially funded by
%  blinded for review.
 NSF Grant \#1908762.
\bibliographystyle{plain}
\bibliography{bibmain}

\begin{thebibliography}{100}

\bibitem{agrawal2018better}
A.~Agrawal and T.~Menzies.
\newblock Is better data better than better data miners?: on the benefits of
  tuning smote for defect prediction.
\newblock In {\em IST}. ACM, 2018.

\bibitem{agrawal2018wrong}
Amritanshu Agrawal, Wei Fu, and Tim Menzies.
\newblock What is wrong with topic modeling? and how to fix it using
  search-based software engineering.
\newblock {\em Information and Software Technology}, 98:74--88, 2018.

\bibitem{agarwal17}
Amritanshu Agrawal and Tim Menzies.
\newblock "better data" is better than "better data miners" (benefits of tuning
  {SMOTE} for defect prediction).
\newblock {\em CoRR}, abs/1705.03697, 2017.

\bibitem{agrawal2018we}
Amritanshu Agrawal, Akond Rahman, Rahul Krishna, Alexander Sobran, and Tim
  Menzies.
\newblock We don't need another hero? the impact of" heroes" on software
  development.
\newblock In {\em Proceedings of the 40th International Conference on Software
  Engineering: Software Engineering in Practice}, pages 245--253, 2018.

\bibitem{arcuri2011practical}
Andrea Arcuri and Lionel Briand.
\newblock A practical guide for using statistical tests to assess randomized
  algorithms in software engineering.
\newblock In {\em Software Engineering (ICSE), 2011 33rd International
  Conference on}, pages 1--10. IEEE, 2011.

\bibitem{Arisholm:2006}
E.~Arisholm and L.~C Briand.
\newblock Predicting fault-prone components in a java legacy system.
\newblock In {\em ESEM}. ACM, 2006.

\bibitem{arisholm2010systematic}
Erik Arisholm, Lionel~C Briand, and Eivind~B Johannessen.
\newblock A systematic and comprehensive investigation of methods to build and
  evaluate fault prediction models.
\newblock {\em Journal of Systems and Software}, 83(1):2--17, 2010.

\bibitem{basili1996validation}
Victor~R Basili, Lionel~C Briand, and Walc{\'e}lio~L Melo.
\newblock A validation of object-oriented design metrics as quality indicators.
\newblock {\em Software Engineering, IEEE Transactions on}, 22(10):751--761,
  1996.

\bibitem{bird09reliabity}
C.~{Bird}, N.~{Nagappan}, H.~{Gall}, B.~{Murphy}, and P.~{Devanbu}.
\newblock Putting it all together: Using socio-technical networks to predict
  failures.
\newblock In {\em ISSRE}, 2009.

\bibitem{bird2009does}
Christian Bird, Nachiappan Nagappan, Premkumar Devanbu, Harald Gall, and
  Brendan Murphy.
\newblock Does distributed development affect software quality? an empirical
  case study of windows vista.
\newblock In {\em 2009 IEEE 31st International Conference on Software
  Engineering}, pages 518--528. IEEE, 2009.

\bibitem{bird2011don}
Christian Bird, Nachiappan Nagappan, Brendan Murphy, Harald Gall, and Premkumar
  Devanbu.
\newblock Don't touch my code! examining the effects of ownership on software
  quality.
\newblock In {\em Proceedings of the 19th ACM SIGSOFT symposium and the 13th
  European conference on Foundations of software engineering}, pages 4--14,
  2011.

\bibitem{briand1993developing}
Lionel~C Briand, VR~Brasili, and Christopher~J Hetmanski.
\newblock Developing interpretable models with optimized set reduction for
  identifying high-risk software components.
\newblock {\em IEEE Transactions on Software Engineering}, 19(11):1028--1044,
  1993.

\bibitem{cao2018improved}
Yang Cao, Zhiming Ding, Fei Xue, and Xiaotao Rong.
\newblock An improved twin support vector machine based on multi-objective
  cuckoo search for software defect prediction.
\newblock {\em International Journal of Bio-Inspired Computation},
  11(4):282--291, 2018.

\bibitem{chawla2002smote}
Nitesh~V Chawla, Kevin~W Bowyer, Lawrence~O Hall, and W~Philip Kegelmeyer.
\newblock Smote: synthetic minority over-sampling technique.
\newblock {\em Journal of artificial intelligence research}, 16:321--357, 2002.

\bibitem{chen2018applications}
Di~Chen, Wei Fu, Rahul Krishna, and Tim Menzies.
\newblock Applications of psychological science for actionable analytics.
\newblock {\em FSE'19}, 2018.

\bibitem{csm19}
Di~Chen, Kathryn~T. Stolee, and Tim Menzies.
\newblock Replication can improve prior results: A github study of pull request
  acceptance.
\newblock In {\em Proceedings of the 27th International Conference on Program
  Comprehension}, ICPC '19, page 179–190. IEEE Press, 2019.

\bibitem{choudhary2018empirical}
Garvit~Rajesh Choudhary, Sandeep Kumar, Kuldeep Kumar, Alok Mishra, and Cagatay
  Catal.
\newblock Empirical analysis of change metrics for software fault prediction.
\newblock {\em Computers \& Electrical Engineering}, 67:15--24, 2018.

\bibitem{d2010extensive}
Marco D'Ambros, Michele Lanza, and Romain Robbes.
\newblock An extensive comparison of bug prediction approaches.
\newblock In {\em 2010 7th IEEE Working Conference on Mining Software
  Repositories (MSR 2010)}, pages 31--41. IEEE, 2010.

\bibitem{efron94}
Bradley Efron and Robert~J Tibshirani.
\newblock {\em An introduction to the bootstrap}.
\newblock Mono. Stat. Appl. Probab. London, 1994.

\bibitem{fenton2000software}
Norman~E Fenton and Martin Neil.
\newblock Software metrics: roadmap.
\newblock In {\em Proceedings of the Conference on the Future of Software
  Engineering}, pages 357--370, 2000.

\bibitem{fu2016tuning}
Wei Fu, Tim Menzies, and Xipeng Shen.
\newblock Tuning for software analytics: Is it really necessary?
\newblock {\em Information and Software Technology}, 76:135--146, 2016.

\bibitem{Gao11}
Kehan Gao, Taghi~M. Khoshgoftaar, Huanjing Wang, and Naeem Seliya.
\newblock Choosing software metrics for defect prediction: an investigation on
  feature selection techniques.
\newblock {\em Software: Practice and Experience}, 41(5):579--606, 2011.

\bibitem{ghotra15}
B.~Ghotra, S.~McIntosh, and A.~E. Hassan.
\newblock Revisiting the impact of classification techniques on the performance
  of defect prediction models.
\newblock In {\em 2015 37th ICSE}, 2015.

\bibitem{ghotra2015revisiting}
Baljinder Ghotra, Shane McIntosh, and Ahmed~E Hassan.
\newblock Revisiting the impact of classification techniques on the performance
  of defect prediction models.
\newblock In {\em 37th ICSE-Volume 1}, pages 789--800. IEEE Press, 2015.

\bibitem{giger2012method}
Emanuel Giger, Marco D'Ambros, Martin Pinzger, and Harald~C Gall.
\newblock Method-level bug prediction.
\newblock In {\em Proceedings of the 2012 ACM-IEEE International Symposium on
  Empirical Software Engineering and Measurement}, pages 171--180. IEEE, 2012.

\bibitem{graves2000predicting}
T.~L Graves, A.~F Karr, J.~S Marron, and H.~Siy.
\newblock Predicting fault incidence using software change history.
\newblock {\em TSE}, 2000.

\bibitem{he2012investigation}
Zhimin He, Fengdi Shu, Ye~Yang, Mingshu Li, and Qing Wang.
\newblock An investigation on the feasibility of cross-project defect
  prediction.
\newblock {\em Automated Software Engineering}, 19(2):167--199, 2012.

\bibitem{Herbsleb14}
James Herbsleb.
\newblock Socio-technical coordination (keynote).
\newblock In {\em Companion Proceedings of the 36th International Conference on
  Software Engineering}, ICSE Companion 2014, page~1, New York, NY, USA, 2014.
  Association for Computing Machinery.

\bibitem{huang2017supervised}
Qiao Huang, Xin Xia, and David Lo.
\newblock Supervised vs unsupervised models: A holistic look at effort-aware
  just-in-time defect prediction.
\newblock In {\em Software Maintenance and Evolution (ICSME), 2017 IEEE
  International Conference on}, pages 159--170. IEEE, 2017.

\bibitem{ibrahim2017software}
Dyana~Rashid Ibrahim, Rawan Ghnemat, and Amjad Hudaib.
\newblock Software defect prediction using feature selection and random forest
  algorithm.
\newblock In {\em 2017 International Conference on New Trends in Computing
  Sciences (ICTCS)}, pages 252--257. IEEE, 2017.

\bibitem{jacob2015improved}
Shomona~Gracia Jacob et~al.
\newblock Improved random forest algorithm for software defect prediction
  through data mining techniques.
\newblock {\em International Journal of Computer Applications}, 117(23), 2015.

\bibitem{perils}
Eirini Kalliamvakou, Georgios Gousios, Kelly Blincoe, Leif Singer, Daniel~M.
  German, and Daniela Damian.
\newblock The promises and perils of mining github.
\newblock In {\em Proceedings of the 11th Working Conference on Mining Software
  Repositories}, MSR 2014, pages 92--101, New York, NY, USA, 2014. ACM.

\bibitem{Kamei10}
Y.~{Kamei}, S.~{Matsumoto}, A.~{Monden}, K.~{Matsumoto}, B.~{Adams}, and A.~E.
  {Hassan}.
\newblock Revisiting common bug prediction findings using effort-aware models.
\newblock In {\em 2010 IEEE International Conference on Software Maintenance},
  pages 1--10, 2010.

\bibitem{kamei2010revisiting}
Yasutaka Kamei, Shinsuke Matsumoto, Akito Monden, Ken-ichi Matsumoto, Bram
  Adams, and Ahmed~E Hassan.
\newblock Revisiting common bug prediction findings using effort-aware models.
\newblock In {\em 2010 IEEE International Conference on Software Maintenance},
  pages 1--10. IEEE, 2010.

\bibitem{kamei2007effects}
Yasutaka Kamei, Akito Monden, Shinsuke Matsumoto, Takeshi Kakimoto, and
  Ken-ichi Matsumoto.
\newblock The effects of over and under sampling on fault-prone module
  detection.
\newblock In {\em First International Symposium on Empirical Software
  Engineering and Measurement (ESEM 2007)}, pages 196--204. IEEE, 2007.

\bibitem{kamei2012large}
Yasutaka Kamei, Emad Shihab, Bram Adams, Ahmed~E Hassan, Audris Mockus, Anand
  Sinha, and Naoyasu Ubayashi.
\newblock A large-scale empirical study of just-in-time quality assurance.
\newblock {\em IEEE Transactions on Software Engineering}, 39(6):757--773,
  2012.

\bibitem{kochhar2016practitioners}
Pavneet~Singh Kochhar, Xin Xia, David Lo, and Shanping Li.
\newblock Practitioners' expectations on automated fault localization.
\newblock In {\em Proceedings of the 25th International Symposium on Software
  Testing and Analysis}, pages 165--176. ACM, 2016.

\bibitem{kondo2020impact}
Masanari Kondo, Daniel~M German, Osamu Mizuno, and Eun-Hye Choi.
\newblock The impact of context metrics on just-in-time defect prediction.
\newblock {\em Empirical Software Engineering}, 25(1):890--939, 2020.

\bibitem{krishna2018bellwethers}
Rahul Krishna and Tim Menzies.
\newblock Bellwethers: A baseline method for transfer learning.
\newblock {\em IEEE Transactions on Software Engineering}, 2018.

\bibitem{li2018progress}
Zhiqiang Li, Xiao-Yuan Jing, and Xiaoke Zhu.
\newblock Progress on approaches to software defect prediction.
\newblock {\em IET Software}, 12(3):161--175, 2018.

\bibitem{Lumpe12}
Markus Lumpe, Rajesh Vasa, Tim Menzies, Rebecca Rush, and Burak Turhan.
\newblock Learning better inspection optimization policies.
\newblock {\em International Journal of Software Engineering and Knowledge
  Engineering}, 22(5):621--644, 8 2012.

\bibitem{madeyski2006external}
Lech Madeyski.
\newblock Is external code quality correlated with programming experience or
  feelgood factor?
\newblock In {\em International Conference on Extreme Programming and Agile
  Processes in Software Engineering}, pages 65--74. Springer, 2006.

\bibitem{madeyski2015process}
Lech Madeyski and Marian Jureczko.
\newblock Which process metrics can significantly improve defect prediction
  models? an empirical study.
\newblock {\em Software Quality Journal}, 23(3):393--422, 2015.

\bibitem{mathew2017trends}
George Mathew, Amritanshu Agrawal, and Tim Menzies.
\newblock Trends in topics at se conferences (1993-2013).
\newblock In {\em 2017 IEEE/ACM 39th International Conference on Software
  Engineering Companion (ICSE-C)}, pages 397--398. IEEE, 2017.

\bibitem{matsumoto2010analysis}
S.~Matsumoto, Y.~Kamei, A.~Monden, K.~Matsumoto, and M.~Nakamura.
\newblock An analysis of developer metrics for fault prediction.
\newblock In {\em 6th PROMISE}, 2010.

\bibitem{menzies07dp}
T.~Menzies, J.~Greenwald, and A.~Frank.
\newblock Data mining static code attributes to learn defect predictors.
\newblock {\em TSE}, 2007.

\bibitem{menzies10dp}
T.~Menzies, Z.~Milton, B.~Turhan, B.~Cukic, Y.~Jiang, and A.~Bener.
\newblock Defect prediction from static code features: Current results,
  limitations, new approaches.
\newblock {\em ASE}, 2010.

\bibitem{menzies2006data}
Tim Menzies, Jeremy Greenwald, and Art Frank.
\newblock Data mining static code attributes to learn defect predictors.
\newblock {\em IEEE transactions on software engineering}, 33(1):2--13, 2006.

\bibitem{menzies2018500+}
Tim Menzies, Suvodeep Majumder, Nikhila Balaji, Katie Brey, and Wei Fu.
\newblock 500+ times faster than deep learning:(a case study exploring faster
  methods for text mining stackoverflow).
\newblock In {\em 2018 IEEE/ACM 15th International Conference on Mining
  Software Repositories (MSR)}, pages 554--563. IEEE, 2018.

\bibitem{menzies2008implications}
Tim Menzies, Burak Turhan, Ayse Bener, Gregory Gay, Bojan Cukic, and Yue Jiang.
\newblock Implications of ceiling effects in defect predictors.
\newblock In {\em Proceedings of the 4th international workshop on Predictor
  models in software engineering}, pages 47--54. ACM, 2008.

\bibitem{mittas2013ranking}
Nikolaos Mittas and Lefteris Angelis.
\newblock Ranking and clustering software cost estimation models through a
  multiple comparisons algorithm.
\newblock {\em IEEE Transactions on software engineering}, 39(4):537--551,
  2013.

\bibitem{Moser08}
Raimund Moser, Witold Pedrycz, and Giancarlo Succi.
\newblock A comparative analysis of the efficiency of change metrics and static
  code attributes for defect prediction.
\newblock In {\em Proceedings of the 30th International Conference on Software
  Engineering}, ICSE ’08, page 181–190, New York, NY, USA, 2008.
  Association for Computing Machinery.

\bibitem{moser2008comparative}
Raimund Moser, Witold Pedrycz, and Giancarlo Succi.
\newblock A comparative analysis of the efficiency of change metrics and static
  code attributes for defect prediction.
\newblock In {\em Proceedings of the 30th International Conference on Software
  Engineering}, pages 181--190. ACM, 2008.

\bibitem{curating}
Nuthan Munaiah, Steven Kroh, Craig Cabrey, and Meiyappan Nagappan.
\newblock Curating github for engineered software projects.
\newblock {\em Empirical Software Engineering}, 22(6):3219--3253, Dec 2017.

\bibitem{nagappan2007using}
Nachiappan Nagappan and Thomas Ball.
\newblock Using software dependencies and churn metrics to predict field
  failures: An empirical case study.
\newblock In {\em First International Symposium on Empirical Software
  Engineering and Measurement (ESEM 2007)}, pages 364--373. IEEE, 2007.

\bibitem{nagappan2006mining}
Nachiappan Nagappan, Thomas Ball, and Andreas Zeller.
\newblock Mining metrics to predict component failures.
\newblock In {\em Proceedings of the 28th International Conference on Software
  Engineering}, pages 452--461. ACM, 2006.

\bibitem{nagappan2010change}
Nachiappan Nagappan, Andreas Zeller, Thomas Zimmermann, Kim Herzig, and Brendan
  Murphy.
\newblock Change bursts as defect predictors.
\newblock In {\em 2010 IEEE 21st International Symposium on Software
  Reliability Engineering}, pages 309--318. IEEE, 2010.

\bibitem{nam18tse}
J.~{Nam}, W.~{Fu}, S.~{Kim}, T.~{Menzies}, and L.~{Tan}.
\newblock Heterogeneous defect prediction.
\newblock {\em IEEE TSE}, 2018.

\bibitem{nam2013transfer}
Jaechang Nam, Sinno~Jialin Pan, and Sunghun Kim.
\newblock Transfer defect learning.
\newblock In {\em Software Engineering (ICSE), 2013 35th International
  Conference on}, pages 382--391. IEEE, 2013.

\bibitem{nayrolles2018clever}
Mathieu Nayrolles and Abdelwahab Hamou-Lhadj.
\newblock Clever: combining code metrics with clone detection for just-in-time
  fault prevention and resolution in large industrial projects.
\newblock In {\em Proceedings of the 15th International Conference on Mining
  Software Repositories}, pages 153--164, 2018.

\bibitem{onan2016multiobjective}
Aytu{\u{g}} Onan, Serdar Koruko{\u{g}}lu, and Hasan Bulut.
\newblock A multiobjective weighted voting ensemble classifier based on
  differential evolution algorithm for text sentiment classification.
\newblock {\em Expert Systems with Applications}, 62:1--16, 2016.

\bibitem{ostrand04}
Thomas~J. Ostrand, Elaine~J. Weyuker, and Robert~M. Bell.
\newblock Where the bugs are.
\newblock In {\em ISSTA '04: Proceedings of the 2004 ACM SIGSOFT international
  symposium on Software testing and analysis}, pages 86--96, New York, NY, USA,
  2004. ACM.

\bibitem{pan2010domain}
Sinno~Jialin Pan, Ivor~W Tsang, James~T Kwok, and Qiang Yang.
\newblock Domain adaptation via transfer component analysis.
\newblock {\em IEEE Transactions on Neural Networks}, 22(2):199--210, 2010.

\bibitem{parnin2011automated}
Chris Parnin and Alessandro Orso.
\newblock Are automated debugging techniques actually helping programmers?
\newblock In {\em Proceedings of the 2011 international symposium on software
  testing and analysis}, pages 199--209. ACM, 2011.

\bibitem{pascarella2019fine}
Luca Pascarella, Fabio Palomba, and Alberto Bacchelli.
\newblock Fine-grained just-in-time defect prediction.
\newblock {\em Journal of Systems and Software}, 150:22--36, 2019.

\bibitem{pascarella2020performance}
Luca Pascarella, Fabio Palomba, and Alberto Bacchelli.
\newblock On the performance of method-level bug prediction: A negative result.
\newblock {\em Journal of Systems and Software}, 161:110493, 2020.

\bibitem{radjenovic2013software}
Danijel Radjenovi{\'c}, Marjan Heri{\v{c}}ko, Richard Torkar, and Ale{\v{s}}
  {\v{Z}}ivkovi{\v{c}}.
\newblock Software fault prediction metrics: A systematic literature review.
\newblock {\em Information and software technology}, 55(8):1397--1418, 2013.

\bibitem{rahman2011ownership}
Foyzur Rahman and Premkumar Devanbu.
\newblock Ownership, experience and defects: a fine-grained study of
  authorship.
\newblock In {\em Proceedings of the 33rd International Conference on Software
  Engineering}, pages 491--500, 2011.

\bibitem{Ra13}
Foyzur Rahman and Premkumar Devanbu.
\newblock How, and why, process metrics are better.
\newblock In {\em Proceedings of the 2013 International Conference on Software
  Engineering}, pages 432--441. IEEE Press, 2013.

\bibitem{rahman2013and}
Foyzur Rahman and Premkumar Devanbu.
\newblock How, and why, process metrics are better.
\newblock In {\em Software Engineering (ICSE), 2013 35th International
  Conference on}, pages 432--441. IEEE, 2013.

\bibitem{Devanbu14}
Foyzur Rahman, Sameer Khatri, Earl~T. Barr, and Premkumar Devanbu.
\newblock Comparing static bug finders and statistical prediction.
\newblock In {\em Proceedings of the 36th International Conference on Software
  Engineering}, ICSE 2014, page 424–434, New York, NY, USA, 2014. Association
  for Computing Machinery.

\bibitem{rahman2014comparing}
Foyzur Rahman, Sameer Khatri, Earl~T Barr, and Premkumar Devanbu.
\newblock Comparing static bug finders and statistical prediction.
\newblock In {\em Proceedings of the 36th International Conference on Software
  Engineering}, pages 424--434. ACM, 2014.

\bibitem{rahman2013sample}
Foyzur Rahman, Daryl Posnett, Israel Herraiz, and Premkumar Devanbu.
\newblock Sample size vs. bias in defect prediction.
\newblock In {\em Proceedings of the 2013 9th joint meeting on foundations of
  software engineering}, pages 147--157, 2013.

\bibitem{rahman2011bugcache}
Foyzur Rahman, Daryl Posnett, Abram Hindle, Earl Barr, and Premkumar Devanbu.
\newblock Bugcache for inspections: hit or miss?
\newblock In {\em Proceedings of the 19th ACM SIGSOFT symposium and the 13th
  European conference on Foundations of software engineering}, pages 322--331,
  2011.

\bibitem{commitguru}
C.~Rosen, B.~Grawi, and E.~Shihab.
\newblock Commit guru: Analytics and risk prediction of software commits.
\newblock ESEC/FSE 2015, 2015.

\bibitem{rosen2015commit}
Christoffer Rosen, Ben Grawi, and Emad Shihab.
\newblock Commit guru: analytics and risk prediction of software commits.
\newblock In {\em Proceedings of the 2015 10th Joint Meeting on Foundations of
  Software Engineering}, pages 966--969. ACM, 2015.

\bibitem{ryu2016value}
Duksan Ryu, Okjoo Choi, and Jongmoon Baik.
\newblock Value-cognitive boosting with a support vector machine for
  cross-project defect prediction.
\newblock {\em Empirical Software Engineering}, 21(1):43--71, 2016.

\bibitem{seiffert2014empirical}
Chris Seiffert, Taghi~M Khoshgoftaar, Jason Van~Hulse, and Andres Folleco.
\newblock An empirical study of the classification performance of learners on
  imbalanced and noisy software quality data.
\newblock {\em Information Sciences}, 259:571--595, 2014.

\bibitem{seliya2010predicting}
Naeem Seliya, Taghi~M Khoshgoftaar, and Jason Van~Hulse.
\newblock Predicting faults in high assurance software.
\newblock In {\em 2010 IEEE 12th International Symposium on High Assurance
  Systems Engineering}, pages 26--34. IEEE, 2010.

\bibitem{Shin2013}
Y.~Shin and L.~Williams.
\newblock Can traditional fault prediction models be used for vulnerability
  prediction?
\newblock {\em EMSE}, 2013.

\bibitem{storn1997differential}
R.~Storn and K.~Price.
\newblock Differential evolution--a simple and efficient heuristic for global
  optimization over continuous spaces.
\newblock {\em Journal of global optimization}, 11(4):341--359, 1997.

\bibitem{subramanyam2003empirical}
Ramanath Subramanyam and Mayuram~S. Krishnan.
\newblock Empirical analysis of ck metrics for object-oriented design
  complexity: Implications for software defects.
\newblock {\em IEEE Transactions on software engineering}, 29(4):297--310,
  2003.

\bibitem{sun2012using}
Zhongbin Sun, Qinbao Song, and Xiaoyan Zhu.
\newblock Using coding-based ensemble learning to improve software defect
  prediction.
\newblock {\em IEEE Transactions on Systems, Man, and Cybernetics, Part C
  (Applications and Reviews)}, 42(6):1806--1817, 2012.

\bibitem{Tantithamthavorn18}
C.~Tantithamthavorn, S.~McIntosh, A.~E. Hassan, and K.~Matsumoto.
\newblock The impact of automated parameter optimization on defect prediction
  models.
\newblock {\em IEEE Transactions on Software Engineering}, pages 1--1, 2018.

\bibitem{tantithamthavorn2015impact}
Chakkrit Tantithamthavorn, Shane McIntosh, Ahmed~E Hassan, Akinori Ihara, and
  Kenichi Matsumoto.
\newblock The impact of mislabelling on the performance and interpretation of
  defect prediction models.
\newblock In {\em 2015 IEEE/ACM 37th IEEE International Conference on Software
  Engineering}, volume~1, pages 812--823. IEEE, 2015.

\bibitem{tantithamthavorn2016automated}
Chakkrit Tantithamthavorn, Shane McIntosh, Ahmed~E Hassan, and Kenichi
  Matsumoto.
\newblock Automated parameter optimization of classification techniques for
  defect prediction models.
\newblock In {\em ICSE 2016}, pages 321--332. ACM, 2016.

\bibitem{tantithamthavorn2018impact}
Chakkrit Tantithamthavorn, Shane McIntosh, Ahmed~E Hassan, and Kenichi
  Matsumoto.
\newblock The impact of automated parameter optimization on defect prediction
  models.
\newblock {\em IEEE Transactions on Software Engineering}, 45(7):683--711,
  2018.

\bibitem{tomar2015comparison}
Divya Tomar and Sonali Agarwal.
\newblock A comparison on multi-class classification methods based on least
  squares twin support vector machine.
\newblock {\em Knowledge-Based Systems}, 81:131--147, 2015.

\bibitem{Tu18Tuning}
Huy Tu and Vivek Nair.
\newblock While tuning is good, no tuner is best.
\newblock In {\em FSE SWAN}, 2018.

\bibitem{Tu20_emblem}
Huy Tu, Zhe Yu, and Tim Menzies.
\newblock Better data labelling with emblem (and how that impacts defect
  prediction).
\newblock {\em IEEE Transactions on Software Engineering}, 2020.

\bibitem{turhan2009relative}
Burak Turhan, Tim Menzies, Ay{\c{s}}e~B Bener, and Justin Di~Stefano.
\newblock On the relative value of cross-company and within-company data for
  defect prediction.
\newblock {\em Empirical Software Engineering}, 14(5):540--578, 2009.

\bibitem{wang2013using}
Shuo Wang and Xin Yao.
\newblock Using class imbalance learning for software defect prediction.
\newblock {\em IEEE Transactions on Reliability}, 62(2):434--443, 2013.

\bibitem{weyuker2008too}
Elaine~J Weyuker, Thomas~J Ostrand, and Robert~M Bell.
\newblock Do too many cooks spoil the broth? using the number of developers to
  enhance defect prediction models.
\newblock {\em Empirical Software Engineering}, 13(5):539--559, 2008.

\bibitem{williams2008szz}
Chadd Williams and Jaime Spacco.
\newblock Szz revisited: verifying when changes induce fixes.
\newblock In {\em Proceedings of the 2008 workshop on Defects in large software
  systems}, pages 32--36. ACM, 2008.

\bibitem{xia2018hyperparameter}
Tianpei Xia, Rahul Krishna, Jianfeng Chen, George Mathew, Xipeng Shen, and Tim
  Menzies.
\newblock Hyperparameter optimization for effort estimation.
\newblock {\em arXiv preprint arXiv:1805.00336}, 2018.

\bibitem{xia2016automated}
Xin Xia, Lingfeng Bao, David Lo, and Shanping Li.
\newblock “automated debugging considered harmful” considered harmful: A
  user study revisiting the usefulness of spectra-based fault localization
  techniques with professionals using real bugs from large systems.
\newblock In {\em 2016 IEEE International Conference on Software Maintenance
  and Evolution (ICSME)}, pages 267--278. IEEE, 2016.

\bibitem{xia2016hydra}
Xin Xia, David Lo, Sinno~Jialin Pan, Nachiappan Nagappan, and Xinyu Wang.
\newblock Hydra: Massively compositional model for cross-project defect
  prediction.
\newblock {\em IEEE Transactions on software Engineering}, 42(10):977--998,
  2016.

\bibitem{xia2016collective}
Xin Xia, David Lo, Xinyu Wang, and Xiaohu Yang.
\newblock Collective personalized change classification with multiobjective
  search.
\newblock {\em IEEE Transactions on Reliability}, 65(4):1810--1829, 2016.

\bibitem{yang2017tlel}
Xinli Yang, David Lo, Xin Xia, and Jianling Sun.
\newblock Tlel: A two-layer ensemble learning approach for just-in-time defect
  prediction.
\newblock {\em Information and Software Technology}, 87:206--220, 2017.

\bibitem{yang2015deep}
Xinli Yang, David Lo, Xin Xia, Yun Zhang, and Jianling Sun.
\newblock Deep learning for just-in-time defect prediction.
\newblock In {\em 2015 IEEE International Conference on Software Quality,
  Reliability and Security}, pages 17--26. IEEE, 2015.

\bibitem{yang2016effort}
Yibiao Yang, Yuming Zhou, Jinping Liu, Yangyang Zhao, Hongmin Lu, Lei Xu,
  Baowen Xu, and Hareton Leung.
\newblock Effort-aware just-in-time defect prediction: simple unsupervised
  models could be better than supervised models.
\newblock In {\em Proceedings of the 2016 24th ACM SIGSOFT International
  Symposium on Foundations of Software Engineering}, pages 157--168. ACM, 2016.

\bibitem{ye2014learning}
Xin Ye, Razvan Bunescu, and Chang Liu.
\newblock Learning to rank relevant files for bug reports using domain
  knowledge.
\newblock In {\em Proceedings of the 22nd ACM SIGSOFT International Symposium
  on Foundations of Software Engineering}, pages 689--699, 2014.

\bibitem{zhang2017data}
Feng Zhang, Iman Keivanloo, and Ying Zou.
\newblock Data transformation in cross-project defect prediction.
\newblock {\em Empirical Software Engineering}, 22(6):3186--3218, 2017.

\bibitem{zhang2016cross}
Feng Zhang, Quan Zheng, Ying Zou, and Ahmed~E Hassan.
\newblock Cross-project defect prediction using a connectivity-based
  unsupervised classifier.
\newblock In {\em 2016 IEEE/ACM 38th International Conference on Software
  Engineering (ICSE)}, pages 309--320. IEEE, 2016.

\bibitem{zhang2009investigation}
Hongyu Zhang.
\newblock An investigation of the relationships between lines of code and
  defects.
\newblock In {\em 2009 IEEE International Conference on Software Maintenance},
  pages 274--283. IEEE, 2009.

\bibitem{zhang2007predicting}
Hongyu Zhang, Xiuzhen Zhang, and Ming Gu.
\newblock Predicting defective software components from code complexity
  measures.
\newblock In {\em 13th Pacific Rim International Symposium on Dependable
  Computing (PRDC 2007)}, pages 93--96. IEEE, 2007.

\bibitem{zhou2006empirical}
Yuming Zhou and Hareton Leung.
\newblock Empirical analysis of object-oriented design metrics for predicting
  high and low severity faults.
\newblock {\em IEEE Transactions on software engineering}, 32(10):771--789,
  2006.

\bibitem{zhou2010ability}
Yuming Zhou, Baowen Xu, and Hareton Leung.
\newblock On the ability of complexity metrics to predict fault-prone classes
  in object-oriented systems.
\newblock {\em Journal of Systems and Software}, 83(4):660--674, 2010.

\bibitem{zimmermann2009cross}
Thomas Zimmermann, Nachiappan Nagappan, Harald Gall, Emanuel Giger, and Brendan
  Murphy.
\newblock Cross-project defect prediction: a large scale experiment on data vs.
  domain vs. process.
\newblock In {\em Proceedings of the the 7th joint meeting of the European
  software engineering conference and the ACM SIGSOFT symposium on The
  foundations of software engineering}, pages 91--100. ACM, 2009.

\bibitem{zimmermann2007predicting}
Thomas Zimmermann, Rahul Premraj, and Andreas Zeller.
\newblock Predicting defects for eclipse.
\newblock In {\em Proceedings of the Third International Workshop on Predictor
  Models in Software Engineering}, page~9. IEEE Computer Society, 2007.

\end{thebibliography}

% \newpage
% \input{response_3.tex}

\end{document}